\documentclass[10pt,a4paper]{article}
\usepackage[utf8]{inputenc}
\usepackage{amsmath}
\usepackage{amsfonts}
\usepackage{amssymb}
\usepackage{hyperref}
\usepackage{graphicx}
\usepackage{float}
\usepackage[left=3cm,right=3cm,top=3cm,bottom=3cm]{geometry}

\usepackage{subcaption}

\usepackage{mathrsfs} 

\usepackage{bm}

\usepackage{natbib}
\newcommand{\biblio}{\bibliography{CovidTestModelling}}

\usepackage{tikz}
\usetikzlibrary{calc}
\usetikzlibrary{arrows.meta}
\pgfmathsetmacro{\CompSize}{1 }
\pgfmathsetmacro{\cd}{2.5}
\definecolor{boxColor}{cmyk}{0.2,0.1,0.2,0}
\definecolor{greyColor}{cmyk}{0.1,0.1,0.1,0}
\definecolor{susColor}{cmyk}{0.1,0.1,0.1,0}
\definecolor{recColor}{cmyk}{0.3,0.1,0.1,0}
\definecolor{infColor}{cmyk}{0.1,0.3,0.1,0}

\usepackage[disable]{todonotes}

\title{What you saw is what you got? - Correcting reported incidence data for testing intensity}
\author{Rasmus Kristoffer Pedersen$^{1,*}$, Christian Berrig$^{1}$, \\ Tamás Tekeli$^{2}$, Gergely Röst$^{2}$, Viggo Andreasen$^{1}$ \\
    \small $^{1}$PandemiX center, Department of Science and Environment, Roskilde University, Denmark  \\
    \small $^{2}$National Laboratory for Health Security, Bolyai Institute, University of Szeged, Szeged 6720, Hungary. \\
    \small $^{*}$Corresponding author, \url{rakrpe@ruc.dk}  \\
}
\date{Pre-print \\ Document compiled on \today}
\begin{document}
\renewcommand{\biblio}{}

\maketitle

\begin{abstract}
        
    During the COVID-19 pandemic, different types of non-pharmaceutical interventions played an important role in the efforts to control outbreaks and to limit the spread of the SARS-CoV-2 virus. In certain countries, large-scale voluntary testing of non-symptomatic individuals was done, with the aim of identifying asymptomatic and pre-symptomatic infections as well as gauging the prevalence in the general population. 



    In this work, we present a mathematical model, used to investigate the dynamics of both observed and unobserved infections as a function of the rate of voluntary testing. 

    The model indicate that increasing the rate of testing causes the observed prevalence to increase, despite a decrease in the true prevalence. 
    For large testing rates, the observed prevalence also decrease. 
    The non-monotonicity of observed prevalence explains some of the discrepancies seen when comparing uncorrected case-counts between countries. An example of such discrepancy is the COVID-19 epidemics observed in Denmark and Hungary during winter 2020/2021, for which the reported case-counts were comparable, but the true prevalence were very different. 
    The model provides a quantitative measure for the ascertainment rate between observed and true incidence, allowing for test-intensity correction of incidence data. 




    By comparing the model to the country-wide epidemic of the Omicron variant (BA.1 and BA.2) in Denmark during the winter 2021/2022, we find a good agreement between the cumulative incidence as estimated by the model and as suggested by serology-studies. 

    While the model does not capture the full complexity of epidemic outbreaks and the effect of different interventions, it provides a simple way to correct raw case-counts for differences in voluntary testing, making comparison across international borders and testing behavior possible.
\end{abstract}

\section{Introduction}

    

    In the first years of the COVID-19 pandemic, different efforts of surveillance for COVID-19 cases were employed, both across borders and over time within individual countries. 
    A range of non-pharmaceutical interventions (NPIs) played an integral role in epidemic control, with Polymerase Chain-Reaction (PCR) testing individuals suspected for SARS-CoV-2 infection reported to be one of the most effective interventions \citep{rannan-eliya_increased_2021}. In some countries, large scale testing of asymptomatic members of the general population was also employed \citep{frnda_pilot_2021}\todo{Include some reference to the Danish situation}.
    As such, testing was not only used as a diagnostic tool to confirm symptomatic infections, but also to identify asymptomatic infections. 
    In Denmark, very high rates of testing was achieved through a scaling up of
    PCR-testing capacity throughout the pandemic, combined with availability of fast Lateral Flow Tests (LFT), both offered free-of-charge at public test-centers. 
    The advantage of LFT-testing were twofold when compared with PCR-testing: 
    while compromising a small amount of precision in test-accuracy (sensitivity and specificity)
    , it is possible to cover a large population with lower latency from test to result \citep{leber_comparing_2021}\todo{Include some more references for this point}.
    This makes LFT-testing of the general population tractable as a mitigation strategy \citep{mina_rethinking_2020,larremore_test_2020}. 
    As an incentive for testing, ``Corona-passports'' were introduced in Denmark in 2021, making a negative test (or proof of vaccination) necessary for participating in many aspects of both public and private life (restaurant visits, large public/private events, etc.).
    As a consequence, there were periods where total tests carried out in Denmark exceeded more than 400,000 daily tests, corresponding to about one test per 15 citizens daily. 
    The public testing regime in combination with something analogue to the Danish Corona-passport, causes a correlation between the social activity and test-frequency, which turns the population-wide testing program into a very effective tool for mitigation, as for even quite weak correlations, the effective reduction of $R_{0}$ is amplified compared to the uncorrelated case \citep{berrig_heterogeneity_2022}.

    As is the nature of recorded data in a demographic context, the records are never perfect. 
    Already early in the COVID-19 pandemic, it was clear that differences in testing behavior affected the apparent epidemic dynamics, necessitating robust methods for correcting incidence data for disparities between the true and the recorded number of cases \cite{carletti_covid-19_2020}. As argued by the same authors, mathematical models of the SIR-type are apt for understanding this disparity. Such approaches have been numerous throughout the pandemic, and while we do not intent to give a complete review of modelling work focusing on testing effort, we briefly mention some relevant work. For clarity, we refer to the ratio between recorded cases and all cases as the ``ascertainment rate'' or $\mathcal{A}$, although different expression has been used by the authors cited.
    \citet{macdonald_modelling_2021} explicitly includes the ascertainment rate as a function of testing rate in a SIR-type model, with clear phenomenological reasoning behind the relationship between the ascertainment rate and testing rate. This enables the authors to fit the model to data for the first pandemic wave of all American states, yielding not only estimates of the changes to the ascertainment rate during a period where testing capacity increased, but also to determine differences between states in terms of outbreak response, the basic reproduction number, and ``lockdown fatigue''.
    Applying a SIR-type model to Italian surveillance data, \citet{marziano_estimating_2023} were able to estimate changes in the ascertainment rate over the first two years of the pandemic, determined as 15\% during the initial wave and around 22\% during later waves. In addition, the authors were also able to estimate infection hospitalization ratios and infection fatality ratios throughout the same period, substantiating the reduction in infection risk with increased rates of vaccination\todo{Do we have more examples we want to include?}.
    A key benefit of high testing capacity is that it allows for tracing of close contacts of identified cases. By tracing and subsequently testing such close contacts, some infected individuals may be identified prior to the onset of infectivity, making it possible for infected individuals to self-quarantine before they would have spread the infection to others. The benefit that such contact-tracing may have for mitigation efforts is potentially very significant, and has consequently been investigated in many modelling papers \citep{heidecke_mathematical_2024,zhang_analysing_2022,sturniolo_testing_2021,barbarossa_fleeing_2021,kretzschmar_impact_2020}
    \todo{Double-check articles are correct one. Add more examples?}
    Other approaches to estimating the ascertainment rate can also be found in the literature. Based on data from the United Kingdom, where high rates of PCR- and LFT-testing were also employed, \citet{colman_ascertainment_2023} uses a statistical approach for estimating an ascertainment rate between 20-40\%, while also being able to determine effect that differences in age and vaccination may have on these estimates. 
    Retrospectively, ascertainment rates can be estimated from serological studies, as well as from the infection-fatality-rate (IFR) and the case-fatality-rate (CFR) \citep{rost_early_2020,horvath_covid-19_2022,quick_regression_2021}.

    Most of the work mentioned has focused on testing as a tool for identifying suspected cases, regardless of whether the suspicion came from symptoms or through contact-tracing. As this was the primary goal of testing in most cases, the high levels of voluntary testing employed in Denmark (as well as a few other countries), means that some of the methods applied elsewhere may not be accurate in the Danish setting. 
    This unique challenges was recognized by Danish health authorities, which in October 2020 presented a statistical approach for correcting case-data for testing rates \citep{statens_serum_institut_ssi_ekspertrapport_2020}, 
    a method which was later revised to more accurately correct data as testing rates increased
    \citep{statens_serum_institut_ssi_test-justerede_2021}
    \todo{Comment from Rasmus: I've been unable to find the report from SSI on the second correction where their parameter was set to 0.3 (instead of 0.7 and 0.55). Perhaps we can try to ask Lasse Engbo and/or Rasmus Skytte if they have a copy of the report (since they wrote it), since it seems to be gone from SSI's website}.
    Previous modelling work has investigated how testing rates may affect the dynamics of an epidemic when some voluntary testing is carried out due to symptomatic infections of other diseases \citep{agarwal_symptom-based_2021}. 
    Voluntary testing may depend on (recorded) incidence, possibly in a way similar to that suggested by \citet{macdonald_modelling_2021}. 
    However, the consequences of distinguishing between testing to confirm infections (or suspected infections) and voluntary testing has not yet been explored in detail through mathematical modelling.

    In the present study, we develop a mathematical model to assess the true incidence of an epidemic disease from recorded incidence data. Particularly, we consider data as recorded through a population-wide testing regime as seen in many European countries during the COVID-19 pandemic. 
    In the model, we distinguish between infections recorded due to (confirmatory) testing of symptomatic individuals and those due to voluntary testing among individuals without symptoms. This allows us to investigate the effect increasing voluntary testing has on both epidemic dynamics and on the proportion of cases recorded. 
    By relating the model to specific epidemic waves of COVID-19, we are able to compare a high-testing scenario with a low-testing scenario and to provide a method for determining true incidence, which can supplement serology studies, both during and following a major epidemic wave.


\section{Model presentation}

    We extend the classic SIR-model \citep{anderson_population_1982} such that we are able to model voluntary testing of the general population. 
    The model is extended with two ``exposed'' stages prior to infectivity as well as a pre-symptomatic stage. 
    Furthermore, asymptomatic infections are considered. 
    We assume that individuals experiencing symptoms are PCR-tested in order to confirm their infections. All other groups are assumed to be tested on a regular basis voluntarily, with a rate per citizen we denote $\tau$.
    Tested individuals from the second exposed population, the pre-symptomatic population and the asymptomatic population move to a ``quarantined'' population.
    Symptomatic individuals are assumed to quarantine as well. For our purposes, we consider both the symptomatic quarantine and the quarantine based on voluntary testing to be perfect, i.e. quarantined individuals do not contribute to new infections. The population that contribute to new infections hence consists of the pre-symptomatic and the asymptomatic populations. 

    We refer to the testing due to symptoms as ``confirmatory'' tests, and unless otherwise noted, ``testing'' refers to the testing effort of the non-symptomatic and susceptible population. As such, the total number of tests carried out is the sum of confirmatory tests and voluntary tests. 
    For purposes of analysis, we distinguish between infectious individuals that were quarantined due to voluntary testing and due to symptoms.  

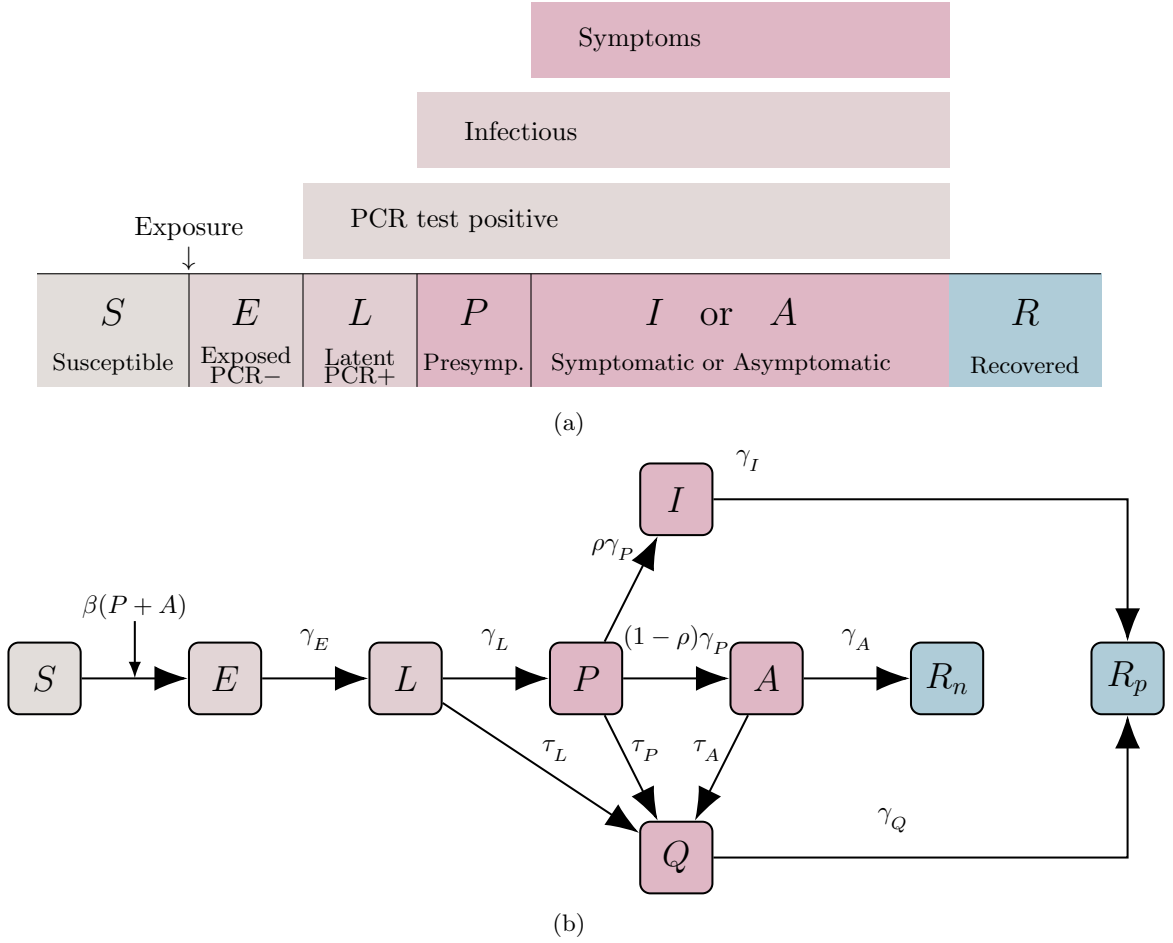
\begin{figure}[ht]
    \begin{subfigure}[t]{\textwidth}
        \centering
            \begin{tikzpicture}

                \draw[draw = none,fill=susColor!90!infColor] (1.5,0.2) rectangle (10,1.2); 
                \draw[draw = none,fill=susColor!70!infColor] (3,1.4) rectangle (10,2.4);
                \draw[draw = none,fill=infColor] (4.5,2.6) rectangle (10,3.6);


                \draw[fill=susColor,draw=none] (-2,-1.5) rectangle (0,0);
                \draw[fill=susColor!80!infColor,draw=none] (0,-1.5) rectangle (1.5,0);
                \draw[fill=susColor!60!infColor,draw=none] (1.5,-1.5) rectangle (3,0);
                \draw[fill=infColor,draw=none] (3,-1.5) rectangle (4.5,0);
                \draw[fill=infColor,draw=none] (4.5,-1.5) rectangle (10,0);
                \draw[fill=recColor,draw=none] (10,-1.5) rectangle (12,0);

                \draw (-2,0) -- (12,0);
                \draw (0,0) -- (0,-1.5);
                \draw (1.5,0) -- (1.5,-1.5);
                \draw (3,0) -- (3,-1.5);
                \draw (4.5,0) -- (4.5,-1.5);

                \node (S) at (-1,-0.5) {\Large $S$};
                \node (S2) at (-1,-1.2) {\small Susceptible};
                \node (E1) at (0.75,-0.5) {\Large $E$};
                \node (E1) at (0.75,-1.1) {\small Exposed};
                \node (E1) at (0.75,-1.35) {\small PCR$-$};
                \node (E2) at (2.25,-0.5) {\Large $L$};
                \node (E2) at (2.25,-1.1) {\small Latent};
                \node (E2) at (2.25,-1.35) {\small PCR+};
                \node (P) at (3.75,-0.5) {\Large $P$};
                \node (P) at (3.75,-1.2) {\small Presymp.};
                \node (IA) at (7,-0.5) {\Large $I$ \; or \; $A$};
                \node (IA) at (7,-1.2) {\small Symptomatic or Asymptomatic};
                \node (R) at (11,-0.5) {\Large $R$};
                \node (R) at (11,-1.2) {\small Recovered};

                \node (exp) at (0,0.2) {$\downarrow$};
                \node (exp2) at (0,0.6) {Exposure};
                \node[anchor=west] (test) at (2,0.7) {PCR test positive};
                \node[anchor=west] (inf) at (3.5,1.9) {Infectious };
                \node[anchor=west] (symp) at (5,3.1) {Symptoms};

            \end{tikzpicture}

        \caption{ }
        \label{fig:DiseaseStages}
    \end{subfigure}
    \begin{subfigure}[t]{\textwidth}
        \centering
        \scalebox{0.95}{
        \begin{tikzpicture}[thick,scale=1, every node/.style={scale=1}]
            \node[rectangle,rounded corners,fill=susColor,draw,minimum size =  \CompSize cm] (S) at (0,0) {\Large $S$};
            \node[rectangle,rounded corners,fill=susColor!80!infColor,draw,minimum size =  \CompSize cm] (E1) at (\cd,0) {\Large $E$};
            \node[rectangle,rounded corners,fill=susColor!60!infColor,draw,minimum size =  \CompSize cm] (E2) at (2*\cd,0) {\Large $L$};
            \node[rectangle,rounded corners,fill=infColor,draw,minimum size =  \CompSize cm] (P) at (3*\cd,0)  {\Large $P$};
            \node[rectangle,rounded corners,fill=infColor,draw,minimum size =  \CompSize cm] (A) at (4*\cd,0) {\Large $A$};
            \node[rectangle,rounded corners,fill=infColor,draw,minimum size =  \CompSize cm] (Q) at (3.5*\cd,-\cd)  {\Large $Q$};
            \node[rectangle,rounded corners,fill=infColor,draw,minimum size =  \CompSize cm] (I) at (3.5*\cd,\cd) {\Large $I$};
            \node[rectangle,rounded corners,fill=recColor,draw,minimum size =  \CompSize cm] (Rn) at (5*\cd,0) {\Large $R_n$};
            \node[rectangle,rounded corners,fill=recColor,draw,minimum size =  \CompSize cm] (Rp) at (6*\cd,0) {\Large $R_p$};

            \draw [-{Latex[scale=2]}]  (S) -- (E1);
            \draw [-{Latex[scale=2]}]  (E1) -- (E2);
            \draw [-{Latex[scale=2]}]  (E2) -- (P);
            \draw [-{Latex[scale=2]}]  (P) -- (A);
            \draw [-{Latex[scale=2]}]  (P) -- (I);
            \draw [-{Latex[scale=2]}]  (E2) -- (Q);
            \draw [-{Latex[scale=2]}]  (P) -- (Q);
            \draw [-{Latex[scale=2]}]  (A) -- (Q);
            \draw [-{Latex[scale=2]}]  (A) -- (Rn);
            \draw [-{Latex[scale=2]}]  (Q) -- ($(Rp)+(0,-\cd)$) -- (Rp);
            \draw [-{Latex[scale=2]}]  (I) -- ($(Rp)+(0,\cd)$) -- (Rp);

            \draw [-{Latex[scale=1]}]  ($(S)!0.5!(E1) + (0,0.8)$) -- ($(S)!0.5!(E1)$);
            \node at ($(S)!0.5!(E1) + (0,1)$) {$\beta (P+A)$};

            \node at ($(E1)!0.5!(E2)+(0,0.5)$) {$\gamma_{{\textstyle\mathstrut}E}$};
            \node at ($(E2)!0.5!(P)+(0,0.5)$) {$\gamma_{{\textstyle\mathstrut}L}$};
            \node at ($(P)!0.5!(A)+(0,0.5)$) {$(1-\rho)\gamma_{{\textstyle\mathstrut}P}$};
            \node at ($(P)!0.5!(I)+(-0.25,0.5)$) {$\rho \gamma_{{\textstyle\mathstrut}P}$};
            \node at ($(E2)!0.5!(Q)+(0.2,0.2)$) {$\tau_{{\textstyle\mathstrut}L}$};
            \node at ($(P)!0.5!(Q)+(0.2,0.2)$) {$\tau_{{\textstyle\mathstrut}P}$};
            \node at ($(A)!0.5!(Q)+(-0.2,0.2)$) {$\tau_{{\textstyle\mathstrut}A}$};
            \node at ($(A)!0.5!(Rn)+(0,0.5)$) {$\gamma_{{\textstyle\mathstrut}A}$};
            \node at ($(I)+(1,0.5)$) {$\gamma_{{\textstyle\mathstrut}I}$};
            \node at ($(Q)+(3,0.5)$) {$\gamma_{{\textstyle\mathstrut}Q}$};

        \end{tikzpicture}
        }
        \caption{}
        \label{fig:ModelFigure}
    \end{subfigure}
        \caption{Model diagrams. (\subref{fig:DiseaseStages}) Diagram of the infection stages considered in the model. Stages $E$, $L$, $P$, $I$ and $A$ are all considered infected. The exposed stage, $E$ is exposed but will not test positive. In the latent stage, $L$, individuals are not yet infectious, but will test positive. Following the pre-symptomatic stage, $P$, a proportion, $\rho$, of individuals develop symptoms, $I$ while others remain asymptomatic, $A$.  (\subref{fig:ModelFigure}) Compartment diagram of the proposed model. Grey symbolizes the susceptible and pre-infectious population, red the infectious population and blue the recovered population. A detailed description of variables and parameters is given in the text and summarized in table \ref{tab:modeldesc}.}
        \label{fig:ModelFiguresSuper}
    \end{figure}

    In figure \ref{fig:DiseaseStages} a diagram of the different stages of an infection is shown for clarification. A compartment diagram of the model is shown in figure \ref{fig:ModelFigure}.

    The model is formulated as a system of ordinary differential equations (ODE's):
    \begin{subequations}
        \begin{align}
            \dot{S} &= - \beta S (P+A) \\ 
            \dot{E} &= \beta S (P+A) - \gamma_E E \\
            \dot{L} &= \gamma_E E - (\gamma_L + \tau_L) L \\
            \dot{P} &= \gamma_L L - (\gamma_P + \tau_P) P\\
            \dot{A} &= \gamma_P (1-\rho) P - (\gamma_A + \tau_A) A \\
            \dot{I} &= \gamma_P \rho P - \gamma_I I \\
            \dot{Q} &= \tau_L L + \tau_P P + \tau_A A - \gamma_Q Q \\
            \dot{R_p} &= \gamma_Q Q + \gamma_I I \label{eq:modeldefinitionRp} \\ 
            \dot{R_n} &= \gamma_A A \label{eq:modeldefinitionRn}
        \end{align}\label{eq:modeldefinition}
        \end{subequations}
        
    with variables scaled by population, such that $S+E+L+P+I+A+Q+R_p+R_n = 1$. 
    A summary of model variables and parameters is given in table \ref{tab:modeldesc}. 
    Although one differential equation can be omitted as the population sums to unity, we analyze the model in its full form as presented above.

    \begin{table}[!ht] \centering
        \caption{Summary of model parameters and variables}\label{tab:modeldesc}
    \begin{tabular}{|r|l||r|l|}
        \hline 
        $S$ & susceptible & $E$ & Exposed (test-negative) \\
        $L$ & Latent (test-positive) & $P$ & Pre-symptomatic \\
        $A$ & Asymptomatic & $I$ & Infectious (symptomatic, quarantined) \\
        $Q$ & Quarantined (due to test) & $R_p$ & Recovered, with positive test \\
        $R_n$ & Recovered, no positive test &&\\
        \hline \hline  
        $\beta$ & Infectivity & $\gamma_y $ & Rate of disease progression from stage $y$ \\
        $\rho $ & Fraction of symptomatic cases & $ \tau_y $ & Rate of (effective) testing at stage $y$ \\ 
        \hline 
    \end{tabular}
    \end{table}

    Note that $\tau_y$ is the effective testing at a given disease stage $y$. We generally assume that actual rates of testing are equal for all non-symptomatic and non-infected disease stages. Hence, the difference between e.g. $\tau_L$ and $\tau_P$ is only an expression of any potential differences in test sensitivity or quality. 

    To avoid tedious notation, we will generally consider the simplifying assumption that $\gamma_E = \gamma_L = \gamma_P = \gamma_A = \gamma$ as well as $\tau_E = \tau_L = \tau_P = \tau_A = \tau$.

    For purposes of evaluating how efficient a given testing intensity, i.e. value of $\tau$, is for identifying infections, we investigate the ascertainment rate, $\mathcal{A}$. We define this as the fraction of cases which are found positive at a given point in time, $t$, out of all cases as the given time:
    \begin{equation}
        \mathcal{A}(t) = \dfrac{R_p(t)}{R_n(t)+R_p(t)}.
    \end{equation}

\section{Results}

In this section, we present the main results of this paper, divided in three main parts: 
Firstly, in section \ref{sec:ModelAnalysis} we describe our model within the framework typically used to analyze SIR-type models and use this framework to give an expression for the ascertainment ratio. 
Secondly, in section \ref{sec:ResultsObservedFinalSize} we present specific observations about the observed epidemic size in comparison to the true epidemic size, based on numerical model investigations.
Finally, we relate the model directly to observed data for the COVID-19 pandemic in section \ref{sec:DataHandlingAndResults}.

\subsection{Model analysis}\label{sec:ModelAnalysis}

    \subsubsection{Formalizing model structure}
    Generalized analysis of the broad family of SIR-type models has been described thoroughly in the literature. Important contributions include the seminal work of \citet{diekmann_definition_1990} as well as \citet{van_den_driessche_reproduction_2002} on the Next Generation Matrix (NGM) method of determining the basic reproduction number, $\mathcal{R}_0$. See \citep{heffernan_perspectives_2005} for a broader review of the importance of $\mathcal{R}_0$. 
    As we will see, the structure provided by the NGM is convenient not only for calculating $\mathcal{R}_0$ and in turn determining the stability of disease free equilibria (DFE), but also for determining the ascertainment ratio, $\mathcal{A}$. For this reason, we first present some initial formalization of our model, which may appear trivial for readers from the field of mathematical epidemiology. 

    Reordering the variables as $x = (E,L,P,A,I,Q,S,R_p,R_n)^T$, all DFE of the system have the form $x_0 = (0,0,0,0,0,0,\sigma,r_p,r_n)^T$, with $1\geq \sigma,r_p,r_n \geq 0$ and $\sigma + r_p + r_n = 1$. We denote the special case where $\sigma=1$ and $r_p = r_n = 0$ as $x_S$.

    For simplicity of notation, we describe the results for the special case where we assume $\gamma_E = \gamma_L = \gamma_P = \gamma_A = \gamma$. 
    In the supplementary material in section \ref{sec:AnalysisGeneral} we show the results of the following analysis for the general form of the model without this assumption. 
    
    Using the notation and nomenclature of \citet{diekmann_construction_2010}, we consider the \textit{infected subsystem} consisting of variables $(E,L,P,A)$ with matrices $\bm{T}$ and $\bm{\Sigma}$ describing the transmission and transition of infections, respectively. In the supplementary material \ref{sec:NGMlarge}, we carry out the analysis for the ``NGM with large domain'', however, here a simplified analysis suffices. 
    Observe that the model contains a single ``state-at-infection'', namely $E$, that is, all newly infected individuals appear in the $E$ compartment. Considering the linearization of the system close to the DFE with $S=1$ and all other variables zero, we define the column-vector $\alpha = (1,0,0,0)'$, and obtain the parts of $\bm{T}$ and $\bm{\Sigma}$ relevant for the present analysis is: $\alpha'\bm{T} = (0,0,\beta, \beta) $ and 
    \begingroup
    \renewcommand*{\arraystretch}{2.2}
    \begin{equation} -\bm{\Sigma}^{-1} \alpha = \begin{pmatrix}
            \dfrac{1}{\gamma} \\
            \dfrac{1}{\gamma+\tau} \\
            \dfrac{\gamma}{(\gamma+\tau)^2} \\
            \dfrac{\gamma^2(1-\rho)}{(\gamma+\tau)^3} \\
        \end{pmatrix}\label{eq:MatrixV}
    \end{equation}
    \endgroup

    The basic reproduction number is given by $\mathcal{R}_0 = -\alpha'\bm{T}\bm{\Sigma}\alpha $, and hence 
    \begin{equation}
        \mathcal{R}_0 =  \beta \left(\dfrac{\gamma}{(\gamma+\tau)^2} + \dfrac{\gamma^2(1-\rho)}{(\gamma+\tau)^3}\right) .
        \label{eq:R0}
    \end{equation}

    \noindent Observe that in the absence of voluntary testing $\mathcal{R}_0|_{\tau=0} = \frac{\beta}{\gamma} \left(2-\rho\right)$.

    Based on derivations similar to those observed in the literature, we show in supplementary material \ref{sec:FinalSizeCalculations} that the epidemic final size, $X = 1-\sigma = r_p + r_n$, for $\mathcal{R}_0 \geq 1$ is a positive and real solution to the expression
    \begin{equation}
        X = 1 - \exp(-\mathcal{R}_0 X)
        \label{eq:FinalSizeExpression}
    \end{equation}

    Note that the theoretical basis of the NGM method discussed here requires the existence of a global asymptotically stable DFE. Under the assumption that immunity wanes at a small but positive rate $\epsilon$, we can introduce a flow from $R_p$ and $R_n$ to $S$, causing the DFE $x_S$ to become globally stable. 
    This changes neither $\bm{T}$ nor $\bm{\Sigma}$, leaving the above calculations identical. Although the definition of $\mathcal{R}_0$ given here is then under valid for a model with waning immunity, numerical investigations confirm that the same expression is valid as a threshold parameter for the (local) stability of all DFE.

    \subsubsection{Ascertainment rate}
    We wish to determine the ascertainment rate, $\mathcal{A}$, that is, the ratio of the number of detected infections to the total number of infections. 
    The observed epidemic final size, i.e. the size of the epidemic as observed only from recorded cases, is hence given by $\mathcal{A} X$ where $X$ is the solution to equation \eqref{eq:FinalSizeExpression}. 
    To simplify the following argument, we calculate determine the ratio between undetected infections and all infections, i.e. $1-\mathcal{A}$.

    Consider a newly infected individual, appearing in state $E$ of the model. 
From a probabilistic perspective of the compartmental model, we can consider the probability that the individual is not detected throughout their infection, i.e. that they end up in the compartment $R_n$. 
    Individuals in compartment $E$ move to $L$ with probability $\frac{\gamma}{\gamma} = 1$. The probability of moving on to $P$ without detection is given by $\frac{\gamma}{\gamma + \tau}$. Similarly, individuals moves from $P$ to $A$ with probability $\frac{\gamma(1-\rho)}{\gamma + \tau}$ and from $A$ to $R_n$ with probability $\frac{\gamma}{\gamma + \tau}$. 
    The probability of any newly infected individual moving to compartment $R_n$ undetected is the product of these, suggesting that 
    \begin{equation}
        1 - \mathcal{A} = \left(\dfrac{\gamma}{\gamma+\tau}\right) \left(\dfrac{\gamma(1-\rho)}{\gamma + \tau}\right) \left(\dfrac{\gamma}{\gamma + \tau} \right)
    \label{eq:AscertaintmentMarkov}
    \end{equation}

    This type of probabilistic argument provides a simple and intuitive way to calculate various properties of a model, and is connected to the NGM \citep[p. 877]{diekmann_construction_2010}.
    We present an alternative approach to the same computation, based on the NGM directly.

    We define an ``entry'' vector, $\alpha$ representing all sources of new infections (in the infected subsystem $(E,L,P,A)$). In our case, this vector is $\alpha = (1,0,0,0)$, as shown above, corresponding to all new infections beginning in compartment $E$ as above. The only exit of the infected subsystem which leads to compartment $R_n$ is from compartment $A$ with rate $\gamma$. We define an ``exit'' vector $\omega = (0,0,0,\gamma)$. Based on the matrix $\bm{\Sigma}^{-1}$ from equation \eqref{eq:MatrixV}, we can calculate
    \begin{equation}
        1 - \mathcal{A} = \omega \bm{\Sigma}^{-1} \alpha^T = \dfrac{\gamma^3(1-\rho)}{(\gamma + \tau)^3}
    \end{equation}
    which is equal to the expression given in equation \eqref{eq:AscertaintmentMarkov}

    For completeness, observe that the method also allows us to compute the ascertainment rate directly. We consider an alternative ``exit'' vector representing individuals that leave the infected subsystem due to detection (or symptom-onset), $\omega_p = (0,\tau,\tau + \gamma \rho , \tau)$. We then the ascertainment rate as
    \begin{equation}
        \mathcal{A} = \omega_p \bm{\Sigma}^{-1} \alpha^T = \dfrac{\tau}{\gamma + \tau} + \dfrac{\gamma ( \tau + \gamma \rho)}{(\gamma + \tau)^2} + \dfrac{\tau \gamma^2 (1-\rho)}{(\gamma+\tau)^3}
    \end{equation}
    Straight-forward calculations show that this expression can be rewritten in the form given in equation \eqref{eq:AscertaintmentMarkov}.

    Note that $\mathcal{A}$ is independent of $\beta$. This is unsurprising, as $\mathcal{A}$ is only a measure of what happens \emph{after} an individual has been infected. 
    In particular, $\mathcal{A}$ is unaffected by model-assumptions regarding the force of infection. In our model the force of infection was defined as $\Lambda = \beta S (P+A)$, hence assuming perfect quarantine of both symptomatic infected and of test-positive quarantined infected. In the absence of this assumption, the force of infection could be written as $\Lambda_{f}=\beta S (P+A+q_I I + q_Q Q)$ with $q_I,q_Q \in [0,1]$ being parameters for the grade of quarantining. The ascertainment rate is however unchanged, as all newly infected individuals still appear in $E$ upon infection, leaving both the Markov chain considerations and the entry vector $\alpha$ unchanged. 

    An interpretation of this method comes from the interpretation of the matrix $\bm{\Sigma}^{-1}$.
    To quote \citep{diekmann_construction_2010}, ``the element $(-\bm{\Sigma}^{-1})_{ij}$ is the expected time that an individual who presently has state $j$ will spend in state $i$ during its entire future `life' (in the epidemiological sense).''
    In the particular cases discussed here, an individual which enters in compartment $E$ thus spends an average duration of $\frac{\gamma^2(1-\rho)}{(\gamma + \tau)^3}$ in compartment $A$. The proportion of individuals from compartment $E$ that end up in compartment $R_n$ is thus the product of average duration they spend in compartment $A$ and the rate from $A$ to $R_n$, namely $\gamma$.

    Our approach also allows for splitting detected cases by sources of detection. Consider the ``exit'' vectors $\omega_{I}= (0,0,\rho \gamma,0)$ and $\omega_{Q}=(0,\tau,\tau,\tau)$, corresponding to individuals that quarantine due to symptoms and due to voluntary testing, respectively. Letting $\mathcal{A}_I$ denote the proportion of infections that are detected due to symptoms and $\mathcal{A}_Q$ denote those detected due to voluntary testing, we obtain

    \begin{align}
        \mathcal{A}_I = \omega_I \bm{\Sigma}^{-1} \alpha^T &=  \dfrac{\rho\gamma^2}{(\gamma+\tau)^2} \\
        \mathcal{A}_Q = \omega_Q \bm{\Sigma}^{-1} \alpha^T &= \dfrac{\tau}{\gamma + \tau} + \dfrac{\tau \gamma}{(\gamma + \tau)^2} + \dfrac{\tau \gamma^2 (1-\rho)}{(\gamma+\tau)^3}
    \end{align}

    Observe that for $\tau=0$, these expressions simply to $\mathcal{A}_I|_{\tau=0} = \rho$ and $\mathcal{A}_Q|_{\tau=0} = 0$, as expected.


    We have here assumed that all disease-progression parameters are equal $\gamma$. This assumption also allows for reducing the model into a simplified form, see supplementary material \ref{sec:ReducedForm}, for which the model dynamics are determined by the parameter $\rho$ as well as the reduced parameters $a=\tau/\gamma$ and $b = \beta/\gamma$. Note that the reduced parameter $a$ corresponds to the number of voluntary tests performed per time relative to the time it takes to progress each step of the infection. As the model considers four steps of disease progression ($E,L,P,$ and either $A$ or $I$), a value of $a=\frac{1}{4}$ corresponds to an average of one test carried out per person over the same span of time as a typical course of infection. In the reduced form, the ascertainment rate can be written as
    \begin{equation}
        \mathcal{A} = 1- \dfrac{1-\rho}{(1+a)^3}
        \label{eq:Ascertainment_ReducedForm_MainText}
    \end{equation}

\subsection{The observed epidemic size for different testing regimes}\label{sec:ResultsObservedFinalSize}



\begin{figure}[ht]
            \centering
            \includegraphics[width= 0.8 \linewidth]{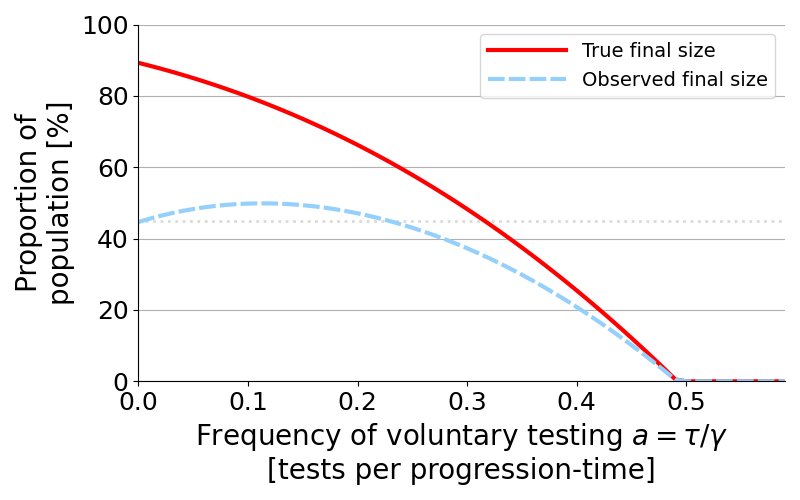}
            \caption{Example illustrating the difference between the true final size and the observed final size. The full red curve shows the true final size, $X$,  as determined using expression \eqref{eq:FinalSizeExpression}, while the dashed blue curve shows the observed final size $X\mathcal{A}$. All disease-progression parameters are $\gamma=1/3$ and $\beta$ has been chosen such that $\mathcal{R}_0=2.5$ for $\tau = 0$. The horizontal axis is given in terms of the ratio between voluntary testing and the rate of disease-progression. A thin gray dotted line shows the observed final size in the absence of testing for comparison.}
            \label{fig:FinalSizeObservedExample}
        \end{figure}

        For a given choice of parameters, we can calculate the (true) epidemic final size, $X$, using equation \eqref{eq:FinalSizeExpression}. Using the ascertainment rate, $\mathcal{A}$, from equation \eqref{eq:AscertaintmentMarkov}, we can also determine the observed final size $X\mathcal{A}$, i.e. the sum of the number of infections identified through symptom-based testing as well as voluntary testing. 
        In the absence of voluntary testing, $\mathcal{A} = \rho$. As voluntary testing increases, the true final size decreases. For some choices of parameters, the observed final size will however first increase with increased testing, before declining with the true final size, see figure \ref{fig:FinalSizeObservedExample}. 
        This leads to a situation were the true final size is reduced by voluntary testing but the observed final size is increased. Furthermore, for a specific frequency of testing, $a=a_0$, there will be an observed final size which is equal the observed final size in the absence of testing, i.e. $(X\mathcal{A})\vert_{a=0} = (X\mathcal{A})\vert_{a=a_0}$.
        In the example shown in figure \ref{fig:FinalSizeObservedExample}, this occurs around $a_0 = 0.23$, at which point the observed final size is $0.45$, equal to the case where $a=0$.
        Note that this observation extends for intermediate levels of voluntary testing, and there exists ranges of values $a_1$ and $a_2$, with $0 < a_1 \leq a_2 < a_0$ such that  $(X\mathcal{A})\vert_{a=a_1} = (X\mathcal{A})\vert_{a=a_2}$. Put differently, a scenario with a low level of voluntary testing, $0 < a_1$, can yield an observed final size corresponding to a higher level of voluntary testing, $a_2 < a_0$.


\begin{figure}[ht]
            \centering
            \includegraphics[width= 0.8 \linewidth]{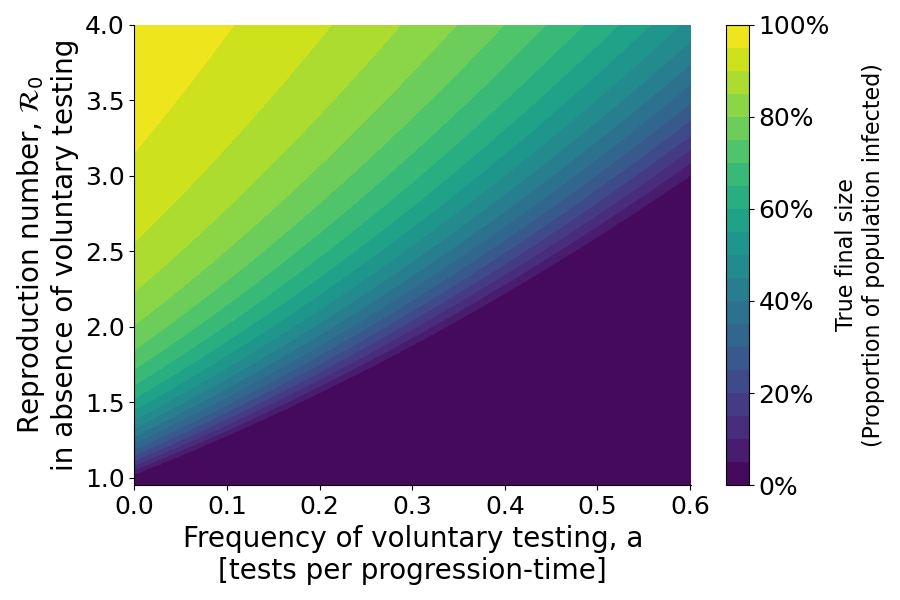}
            \caption{Contour-plot of the true final size for different choices of parameters. The figure illustrates how increased voluntary testing decreases the true epidemic final size. The horizontal axis is given in terms of the ratio between voluntary testing and the disease-progression. The vertical axis is based on a range of value of $\beta$ such that $\mathcal{R}_0$ ranges between $1$ and $4$ when $\tau=0$.}
            \label{fig:FinalSize_TrueContour.png}
        \end{figure}

        In figure \ref{fig:FinalSize_TrueContour.png}, a contour-plot of the true final size is shown for different choices of parameters. Note that the vertical axis is given in terms of $\mathcal{R}_0$. For a fixed value of $\rho$ and $\gamma$, values of $\beta$ can be chosen such that $\mathcal{R}_0$ is within the range shown in the figure. Hence, the contours shown in figure \ref{fig:FinalSize_TrueContour.png} are the same for all choices of $\rho$. 

\begin{figure}[ht]
            \centering
            \includegraphics[width= 0.8 \linewidth]{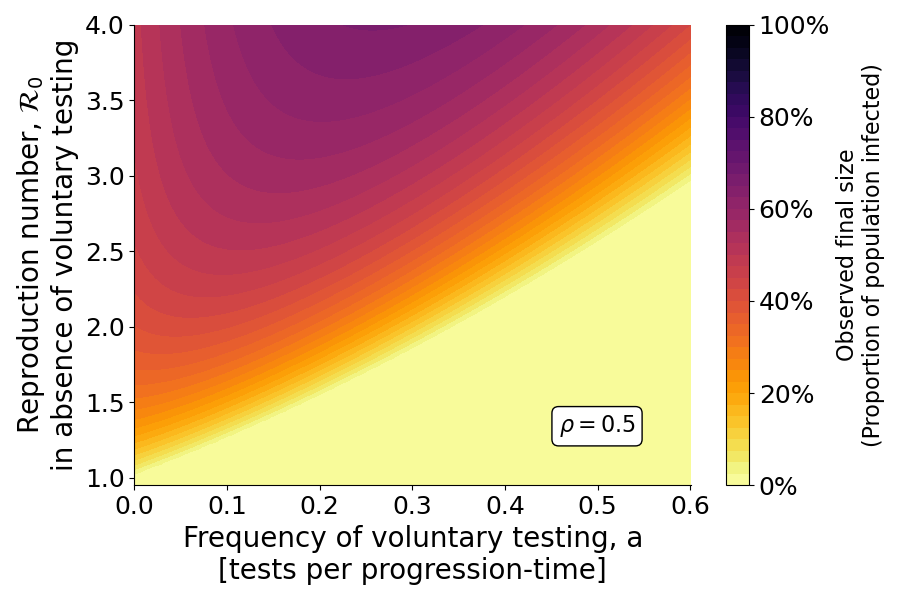}
            \caption{Contour-plot of the observed final size for $\rho=0.5$ and the same range of $\mathcal{R}_0$ and $a$ as in figure \ref{fig:FinalSize_TrueContour.png}. When $a$ is sufficiently large, no epidemic occurs, and the observed final size is $0\%$. For low value of $\mathcal{R}_0$, testing reduces both the true final size (see figure \ref{fig:FinalSize_TrueContour.png}) and the observed final size. For higher values of $\mathcal{R}_0$, the frequency of voluntary testing can lead to an increase in the observed final size, seen in the plot by following a horizontal line from left to right.}
            \label{fig:FinalSize_ObservedContour_50}
        \end{figure}

\begin{figure}
        \begin{subfigure}[b]{0.49\textwidth}
            \centering
            \includegraphics[width=\linewidth]{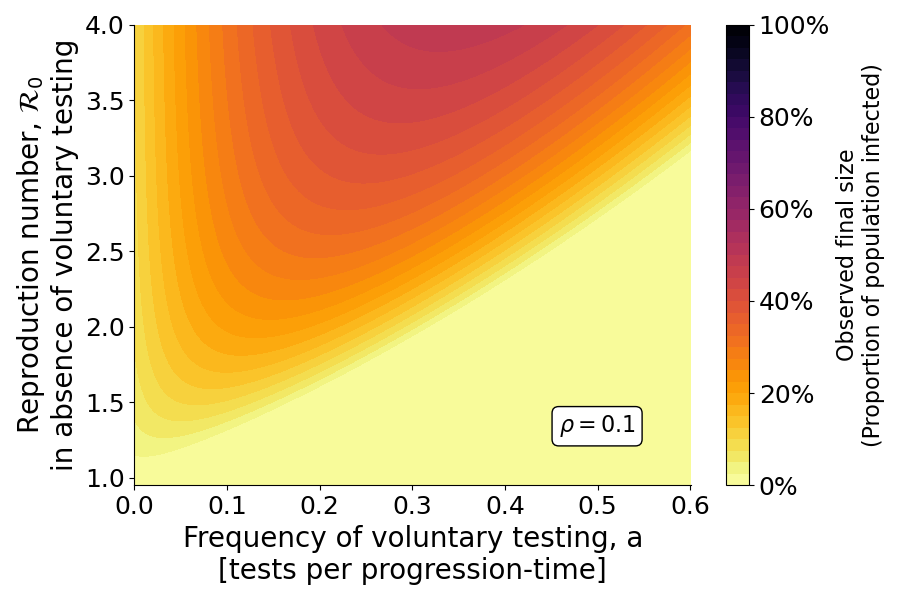} 
            \caption{$\rho = 0.10$ }
            \label{fig:FinalSize_ObservedContour_10}
        \end{subfigure}
        \begin{subfigure}[b]{0.49\textwidth}
            \centering
            \includegraphics[width=\linewidth]{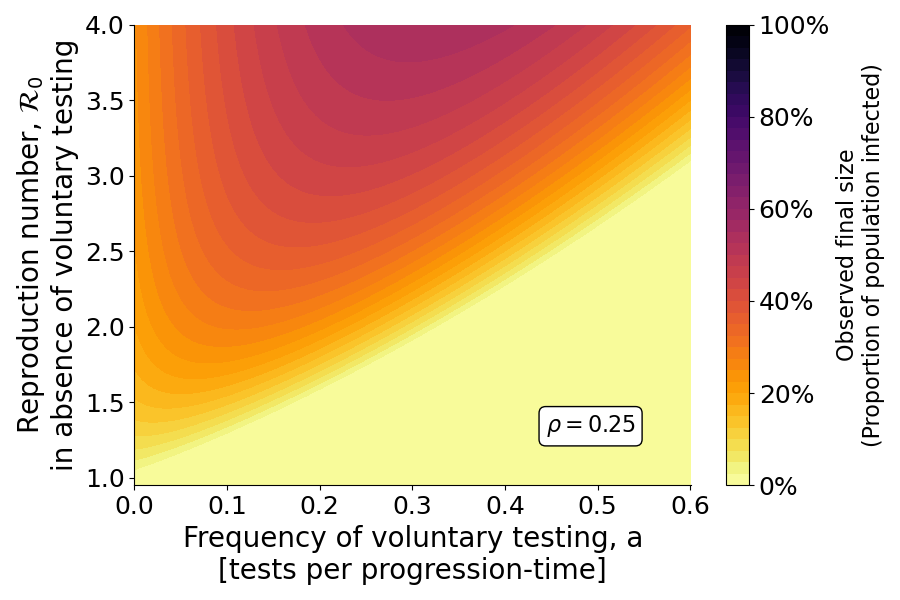}  
            \caption{$\rho = 0.25$ }
            \label{fig:FinalSize_ObservedContour_25}
        \end{subfigure}
        \begin{subfigure}[b]{0.49\textwidth}
            \centering
            \includegraphics[width=\linewidth]{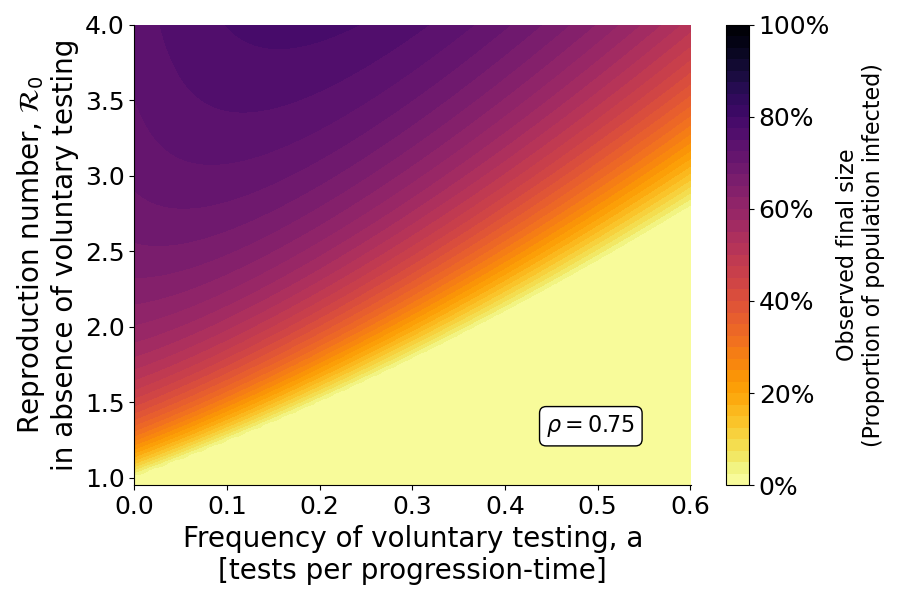}  
            \caption{$\rho = 0.75$ }
            \label{fig:FinalSize_ObservedContour_75}
        \end{subfigure}
        \begin{subfigure}[b]{0.49\textwidth}
            \centering
            \includegraphics[width=\linewidth]{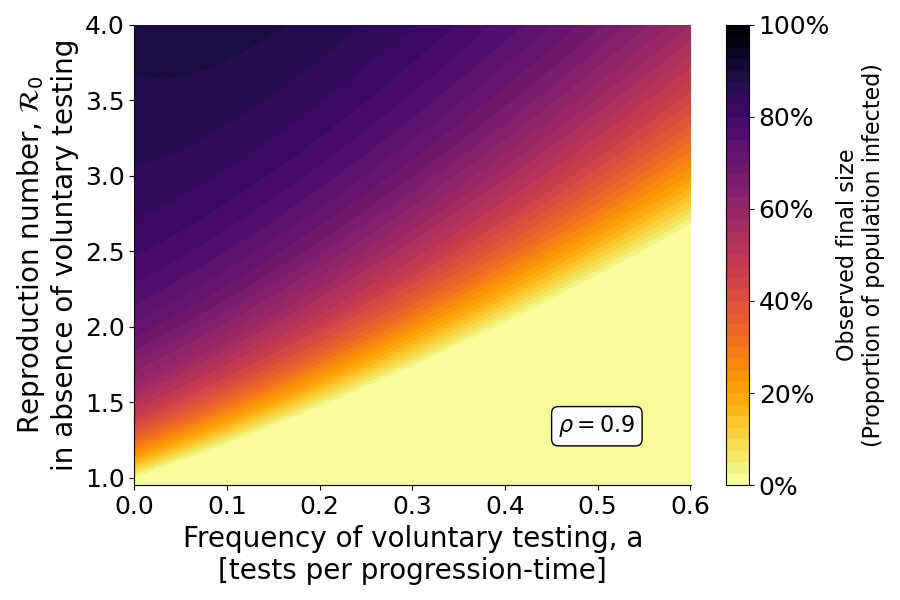}  
            \caption{$\rho = 0.90$ }
            \label{fig:FinalSize_ObservedContour_90}
        \end{subfigure}
            \caption{Contour-plot as in figure \ref{fig:FinalSize_ObservedContour_50} for a range of value of $\rho$. See the figure caption for figure \ref{fig:FinalSize_ObservedContour_50} for details. Note that for $\rho=0.9$, the non-monotonicity only appears for $\mathcal{R}_0 > 3.6$.}
            \label{fig:FinalSize_ObservedContour_Super}
        \end{figure}

        While the relation between the true epidemic size and the frequency of voluntary testing only depends on the compound quantity $\mathcal{R}_0$, the observed final size is also affected by the rate of symptom-based testing, $\rho$. In figure \ref{fig:FinalSize_ObservedContour_50} a contour-plot of the observed final is shown for $\rho=0.5$, while the panels of figure \ref{fig:FinalSize_ObservedContour_Super} show the observed final size for other choices of $\rho$. 
        The figures demonstrate that, as long as $\rho$ is not close to $1$, it is possible to observe scenarios as the one shown in figure \ref{fig:FinalSizeObservedExample}, in which two different choices of frequencies of voluntary testing yields the same observed final size. When $\rho$ is sufficiently low, (e.g. $\rho=0.1$ as in figure \ref{fig:FinalSize_ObservedContour_10}), this may occur for $\mathcal{R}_0$ below $2$ and a relatively low frequency of testing. 

        $X$ decreases monotonically as a function of $a$ and while $\mathcal{A}$ is monotonically increasing and approaches unity asymptotically, the necessary condition for having two equal states is that the derivative of $X\mathcal{A}$ with regard to $a$ evaluated at $a=0$ is positive. This corresponds to the interpretation that for two equal states to be attainable, any increase in voluntary testing from zero must increase the observed final size. The chain-rule yields the expression:
        $$ \dfrac{d(X\mathcal{A})}{da} = X\dfrac{d\mathcal{A}}{da} + \mathcal{A} \dfrac{dX}{da} $$

        Evaluating at $a=0$, we find $\mathcal{A}\vert_{a=0} = \rho$ and $\frac{d\mathcal{A}}{da}\vert_{a=0} = 3-3\rho$. 

\begin{figure}[ht]
            \centering
            \includegraphics[width= 0.8 \linewidth]{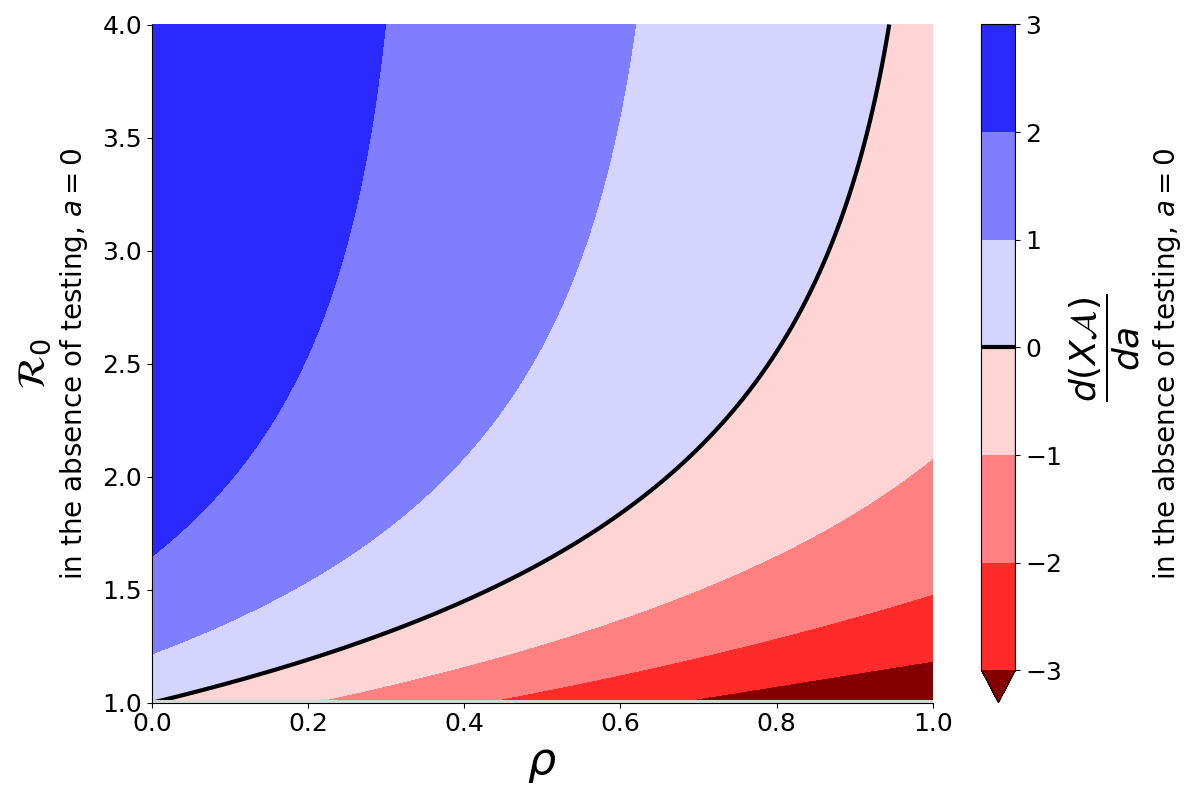}
            \caption{Contour-plot of $\frac{d(X\mathcal{A})}{da}$ evaluated at $a=0$. The black curve shows the contour where the slope is zero. Any set of $\rho$ and $\mathcal{R}_0$ below the curve (shown in red) corresponds to situations where the observed epidemic final size decreases for any increase in voluntary testing above $0$, while points above the curve (shown in blue) exhibits an initial increase in the observed final size when voluntary testing is introduced. Hence, the regime above the black curve allows for the existence of the situation where two epidemics with different voluntary testing can yield observed final sizes that are equal.}
            \label{fig:AscertainmentContour_rhoR0}
        \end{figure}

        With $\mathcal{R}_0\vert_{a=0} = (2-\rho)b$, we determine the $X\vert_{a=0}$ and $\frac{dX}{da}\vert_{a=0}$ numerically for a range of $\rho$ and $\mathcal{R}_0$, and determine $\frac{d(X\mathcal{A})}{da}\vert_{a=0}$. The results are shown in figure \ref{fig:AscertainmentContour_rhoR0}. 
        We observe that for low values of $\rho$, the condition for having two equal states of observed final size occurs for relatively low $\mathcal{R}_0$, while higher values of $\mathcal{R}_0$ is necessary when $\rho$ is higher. For a scenario corresponding to a COVID-19 epidemic with easily available symptomatic testing we assume $\rho=0.5$, suggesting two equal observed final size can occur when $\mathcal{R}_0\geq 1.62$, or, equivalently, that when $\mathcal{R}_0\geq 1.62$, any increase of voluntary testing from zero will result in an increase in the observed size of the epidemic, despite a reduction in infections. 

\clearpage
\pagebreak

\subsection{Data and data-handling}\label{sec:DataHandlingAndResults}



                
                
                
                
                
                





        \subsubsection{Data sources}
        Datasets of international COVID-19 weekly case-data and testing data compiled by ``Our World In Data'' was used in this work \citep{hasell_cross-country_2020,world_health_organization_who_2023}
        For Denmark, additional data was used, all from publicly available sources. In particular, we use daily case-counts supplied by Statens Serum Institut\citep{statens_serum_institut_ssi_filer_2023}. 
        During the Omicron wave in Denmark, additional data was published daily, detailing the current proportion of COVID-19-variants among positive cases\citep{statens_serum_institut_ssi_variant-pcr_2022}.
        Six serology-studies were carried out before and during the Omicron wave, estimating the seroprevalence of the Omicron variant\citep{statens_serum_institut_ssi_seropraevalensundersogelse_2022}.

        \subsubsection{The winter wave of 2020/2021}
        At the end of 2020, many European countries saw a resurgence of COVID-19 cases, in some places coinciding with Christmas celebrations, to which Danish health authorities  responded with a strengthening of NPIs. The rise in cases due to emergence of the Alpha-variant during December 2020 lead to some countries experiencing an epidemic which continued into the first months of 2021.

\begin{figure}[ht]
            \centering
            \includegraphics[width=\linewidth]{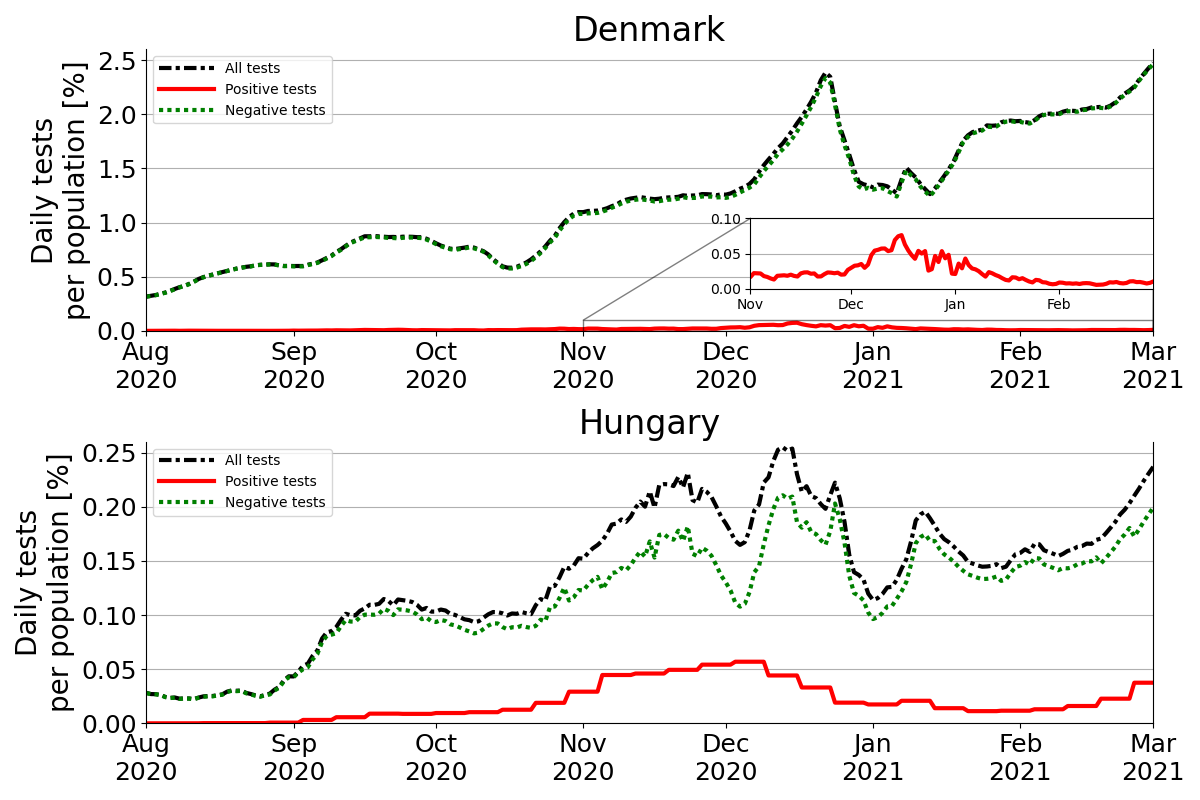}  
            \caption{Data for PCR-tests during the winter epidemic 2020/2021 for Denmark (top panel) and Hungary (bottom panel). Note that different y-axis. The inset in the top panel shows a zoom-in on the period from November 2020 until March 2021. }
            \label{fig:DataFigures_Winter2020}
        \end{figure}

        For the present study, we compare the epidemics experienced in Denmark and in Hungary during the winter of 2020/2021. 
        From data on positive COVID-19 infections, the sum of new cases between October 1$^{st}$ 2020 and February 1$^{st}$ 2021 were comparable, corresponding to about $3\%$ and $3.5\%$ of the population for Denmark and Hungary respectively.
        In Denmark, testing efforts were scaled up during the end of 2020, with voluntary testing becoming a pronounced effort in epidemic mitigation (in addition to localized restrictions to curb the potential of new variants derived from mink, put into effect in parts of the country during October 2020). Throughout the winter, the number of PCR-tests averaged more than one test per hundred citizens daily. In comparison, testing in Hungary was an order of magnitude lower, amounting to between $0.1$ and $0.2$ tests per hundred citizens daily. 
        In figure \ref{fig:DataFigures_Winter2020} time-series data for Denmark and Hungary during the winter 2020/2021 epidemic is shown.

\begin{figure}[ht]
        \begin{subfigure}[t]{0.5\textwidth}
            \centering
            \includegraphics[width=\linewidth]{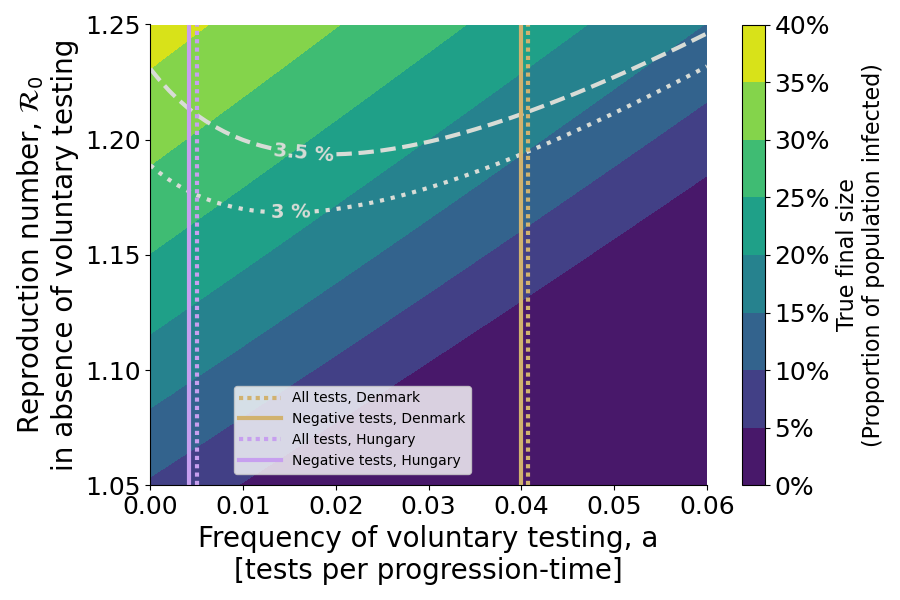}  
            \caption{True final size}
            \label{fig:Contour_DKandHU_Winter2020_True}
        \end{subfigure}
        \begin{subfigure}[t]{0.5\textwidth}
            \centering
            \includegraphics[width=\linewidth]{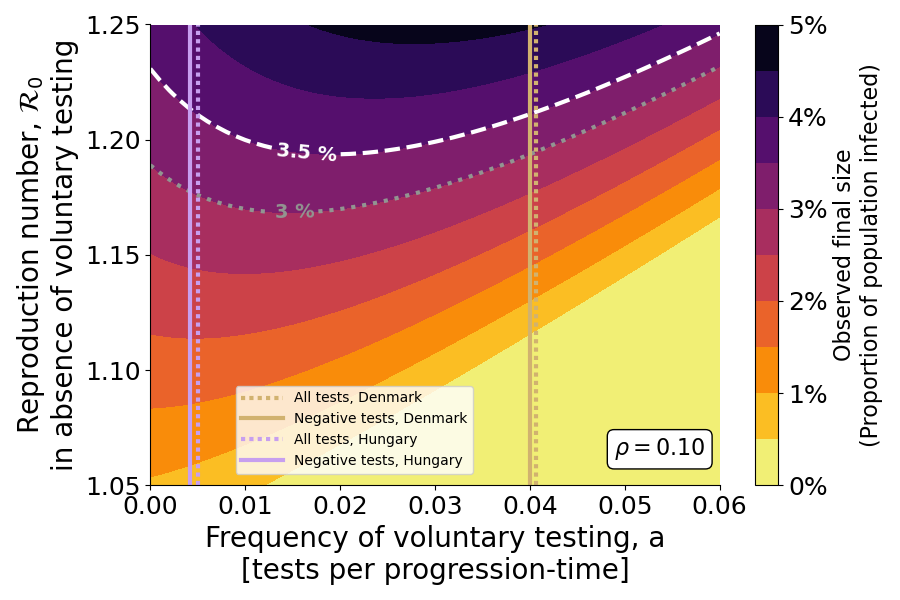}  
            \label{fig:Contour_DKandHU_Winter2020_Observed}
            \caption{Observed final size }
        \end{subfigure}
            \caption{Contour-plot of the true and observed final size for small values of $\mathcal{R}_0$ and $a$ for $\rho=0.1$, with an observed wave of $3\%$ and $3.5\%$ of the population highlighted in both panels. The vertical lines illustrate the average number of tests carried out for Denmark and Hungary in the period between October 1$^{st}$ and February 1$^{st}$. The dotted lines show the average number of test performed per 100 citizens, which for Denmark was $1.356$ and for Hungary was $0.168$. The full lines show the total number of test with all positive tests subtracted, which for Denmark was an average of $0.024$ per day per 100 citizens and $0.029$ for Hungary. }
            \label{fig:Contour_DKandHU_Winter2020}
        \end{figure}

\begin{figure}[ht]
        \begin{subfigure}[t]{0.5\textwidth}
            \centering
            \includegraphics[width=\linewidth]{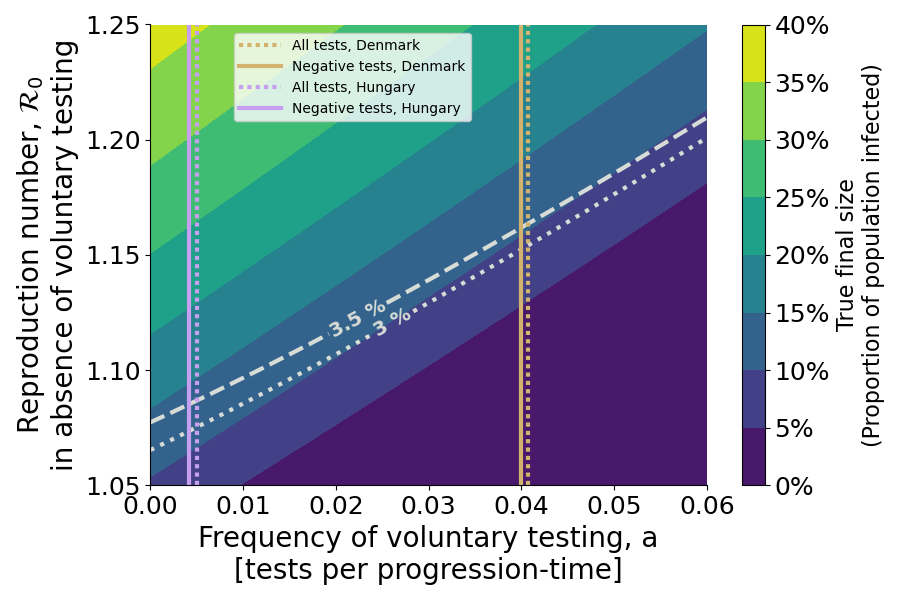}  
            \caption{True final size}
            \label{fig:Contour_DKandHU_Winter2020_True_HigherRho}
        \end{subfigure}
        \begin{subfigure}[t]{0.5\textwidth}
            \centering
            \includegraphics[width=\linewidth]{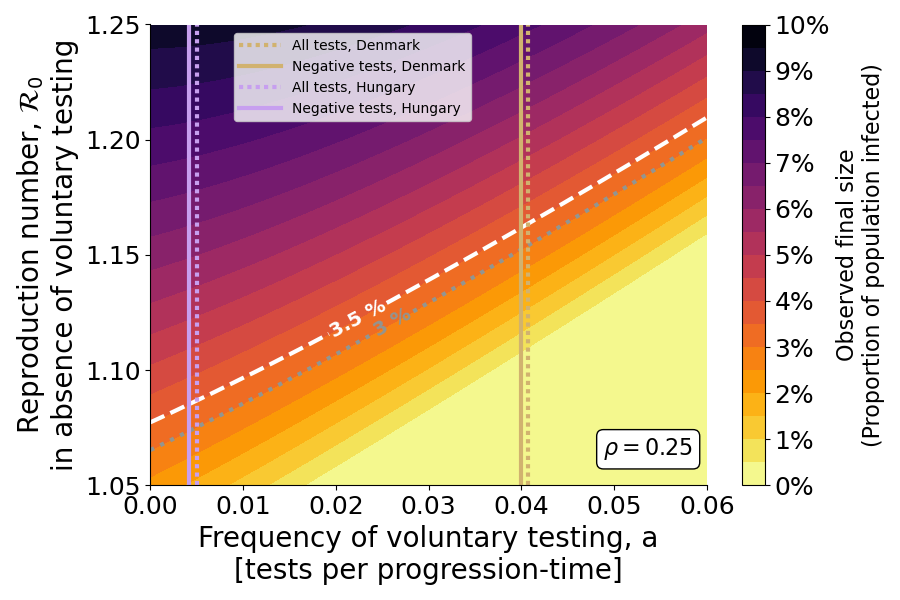}  
            \label{fig:Contour_DKandHU_Winter2020_Observed_HigherRho}
            \caption{Observed final size }
        \end{subfigure}
            \caption{Contour-plot of the true and observed final size for small values of $\mathcal{R}_0$ and $a$ for $\rho=0.25$, with an observed wave of $3\%$ and $3.5\%$ of the population highlighted in both panels. See the figure legend of figure \ref{fig:Contour_DKandHU_Winter2020}.}
            \label{fig:Contour_DKandHU_Winter2020_HigherRho}
        \end{figure}

        Differences in societal response and government instituted NPIs may have influenced transmission dynamics during the winter 2020/2021 epidemic, we assume for simplicity that the epidemic in Denmark and in Hungary had comparable rates of transmission, and can be modelled as an epidemic with similar values of $\mathcal{R}_0$. As discussed in the previous section, epidemics of similar observed size may have very different true size. In figure \ref{fig:Contour_DKandHU_Winter2020} we illustrate how an epidemic wave observed to infect $3\%$ or $3.5\%$ of the population corresponds to both a low and high testing regime when $\rho = 0.1$. 
        For higher values of $\rho$, this correspondence disappears, as shown in figure \ref{fig:Contour_DKandHU_Winter2020_HigherRho} where $\rho = 0.25$. 
        The parameter $\rho$ can be interpreted as the product of the ratio of infected individuals that develop symptoms and the proportion of symptomatic individuals that get their infection confirmed by testing, see supplementary material \ref{sec:ModelExtensionY}. Assuming that half of all infected individuals develop symptoms,  $\rho=0.1$ thus corresponds to one in four having their infection confirmed by testing, while $\rho=0.25$ corresponds to half of symptomatic infections being confirmed. From figures \ref{fig:Contour_DKandHU_Winter2020_True} and \ref{fig:Contour_DKandHU_Winter2020_True_HigherRho}, this assumption about the proportion of symptomatic individuals to get their infection confirmed makes a big difference in the true final size for epidemics with a low observed final size. For the epidemic in Hungary during the winter 2020/2021, $\rho=0.1$ (figure \ref{fig:Contour_DKandHU_Winter2020_True}) implies a true final size above $30\%$ of the population, while $\rho=0.25$ (figure \ref{fig:Contour_DKandHU_Winter2020_True_HigherRho}) implies a true final size just below $15\%$. For the Danish wave, the true final size for $\rho=0.1$ is around $15\%$ while it is $10\%$ for $\rho=0.25$.

        \clearpage

        \subsubsection{The Omicron wave in Denmark}\label{sec:Omicron}

        During the end of 2021 and the beginning of 2022, Denmark experienced a major COVID-19 epidemic, in large part due to the emergence of the Omicron variant in late November 2021, but also due to societal activities related to Christmas and changes in interventions. From November 2021 until April 2022, almost half of all citizens were found positive for COVID-19 infection by PCR-test. 
        Large-scale testing using lateral-flow tests was carried out simultaneously on a voluntary basis, at the order of about 1.5 millions tests weekly, that is, approximately one test per four citizens weekly. Individuals found positive in a lateral-flow test were instructed to get PCR-tested. Voluntary testing with PCR was also available free-of-charge for all citizens throughout the studied period.
        During the Omicron epidemic, a number of serology studies were carried out, estimating that approximately 73\% of all citizens had been infected by the end of April 2022 \citep{statens_serum_institut_ssi_seropraevalensundersogelse_2022}. 
        Throughout December 2021, variant-determination was done for the majority of all positive PCR-tests, resulting in variant-specific incidence data on a daily resolution. 

\begin{figure}
        \begin{subfigure}[b]{\textwidth}
            \centering
            \includegraphics[width=\linewidth]{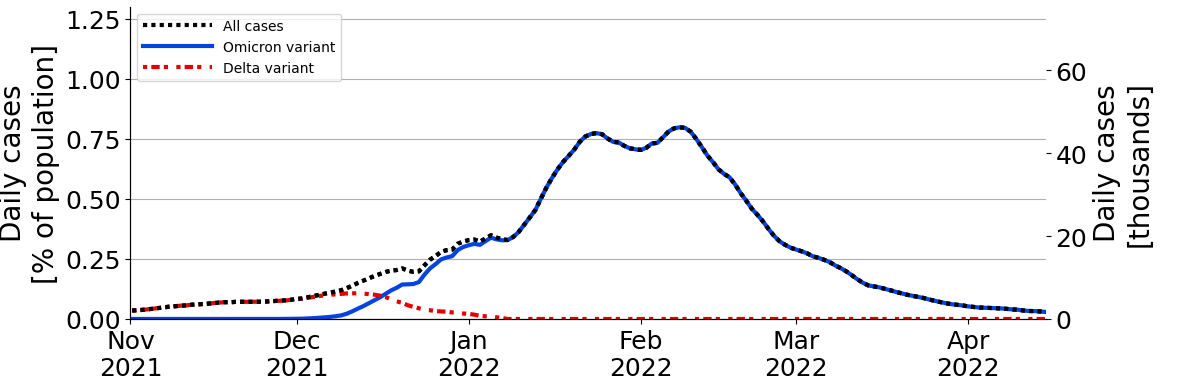}  
            \caption{7-day running mean of positive cases per day. The initial wave of Delta-variant infections is shown in red, while Omicron cases are shown in blue. The sum of all variants are shown in black. }
            \label{fig:DataCases}
        \end{subfigure}
        \begin{subfigure}[b]{\textwidth}
            \centering
            \includegraphics[width=\linewidth]{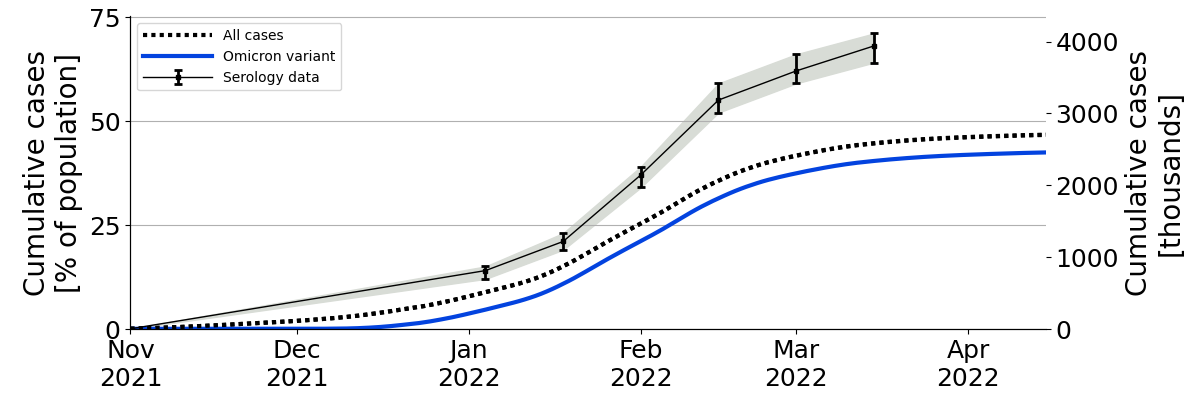}
            \caption{Cumulative cases of detected infections is shown in black, with cumulative Omicron-variant infections shown in blue. The estimate of population-wide serology from \citet{statens_serum_institut_ssi_seropraevalensundersogelse_2022} is shown as black bars with confidence intervals.}
            \label{fig:DataCasesCumulative}
        \end{subfigure}
        \begin{subfigure}[b]{\textwidth}
            \centering
            \includegraphics[width=\linewidth]{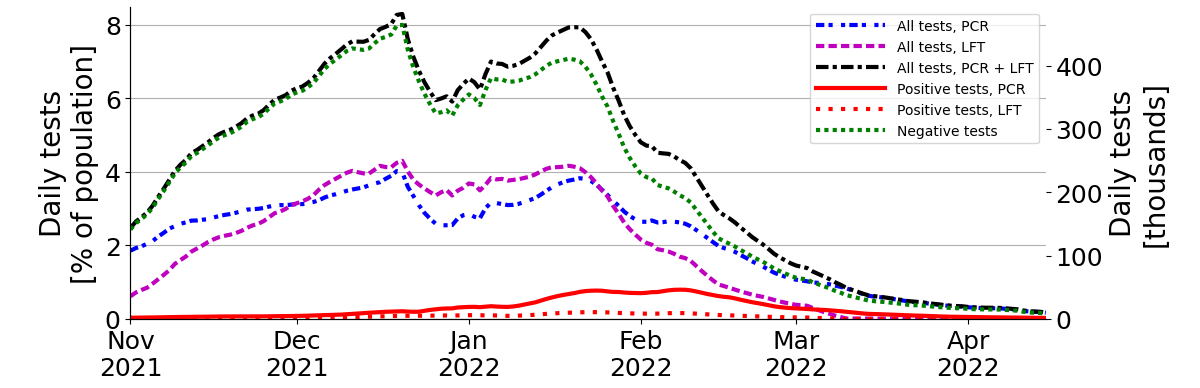} 
            \caption{7-day running mean of daily PCR-tests. The black curve shows the total number of tests carried out. The red curve shows the positive tests, as also shown in panel \ref{fig:DataCases}, with the green curve showing the negative tests calculated as the difference between all tests carried out and the positive tests. The dotted lines show the actual data (right vertical axis), while the full-lines shows the data as a fraction of the subset of the population that had not yet tested positive (left vertical axis), rather than of the full population.}
            \label{fig:DataTests}
        \end{subfigure}
            \caption{}
            \label{fig:DataSuper}
        \end{figure}

        For simplification, we do not distinguish between the two Omicron subtypes ``BA.1'' and ``BA.2'' here. Furthermore, due to the immune-evasion of the Omicron variant, we consider only infections with Omicrons to lead to immunity for future infections (Reinfection-data shows a clear correlation between reinfections and new infections, suggesting that individuals infected by the Delta variant in early December 2021 were susceptible to the Omicron variant during the epidemic). These simplifications allow us to consider the Omicron epidemic as a single wave of a novel virus introduced in a susceptible population. 
        The data is shown in figure \ref{fig:DataSuper}. 

        We assume that most individuals did not get tested again if they had already tested positive during the epidemic. Since about half of the population tested positive throughout the epidemic, it is therefore appropriate to not consider the number of tests carried out daily per citizen, but rather the number of tests carried out per ``citizen that have not previously tested positive''. In figure \ref{fig:DataTests}, we show the number of tests carried out (right axis), as well as the corresponding proportion of the population which had not yet tested positive.

        \begin{figure}[ht]
        \begin{subfigure}[t]{0.5\textwidth}
            \centering
            \includegraphics[width=\linewidth]{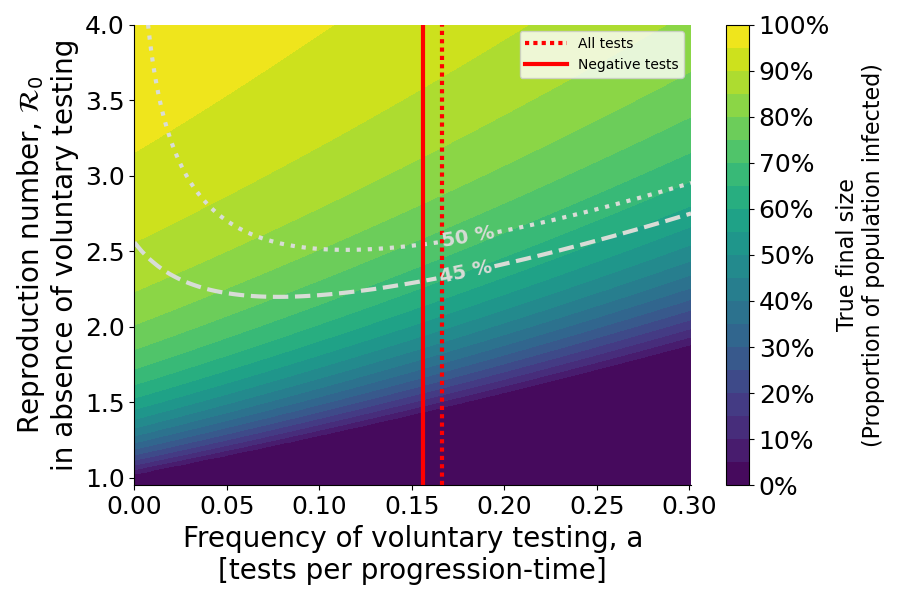}  
            \caption{True final size}
            \label{fig:FinalSize_TrueContour_DK_Omicron}
        \end{subfigure}
        \begin{subfigure}[t]{0.5\textwidth}
            \centering
            \includegraphics[width=\linewidth]{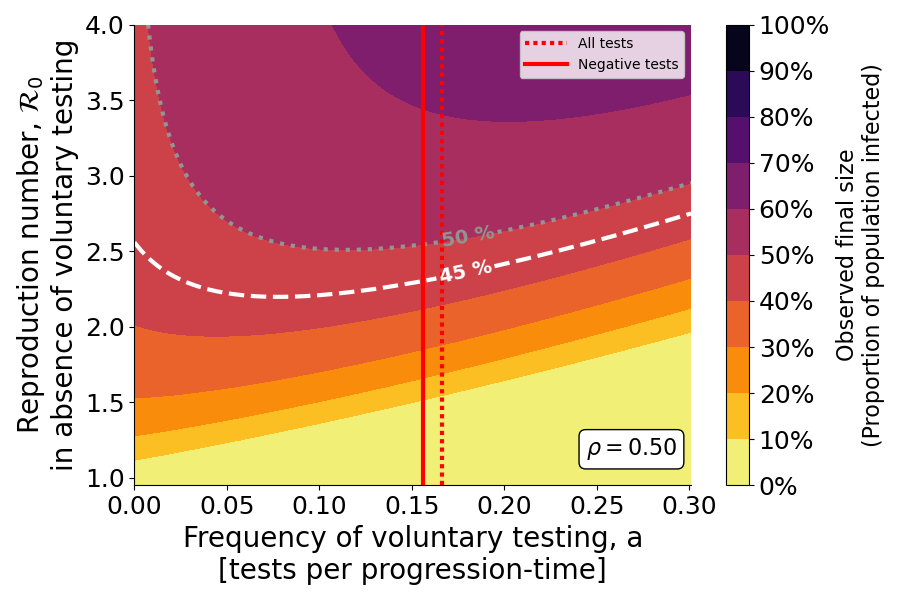}  
            \label{fig:FinalSize_ObservedContour_DK_Omicron}
            \caption{Observed final size}
        \end{subfigure}
            \caption{Contour-plot of the true and observed final size, with an observed wave of $45\%$ and $50\%$ of the population highlighted in both panels. The vertical lines illustrate the average number of tests carried out for Denmark in the period between November 1$^{st}$ 2021 and March 1$^{st}$ 2022. All tests corresponded to an average of $2.853$ test per 100 citizens (dotted line), while subtracting the average of $0.337$ daily positive tests per 100 citizens yield the full line. }
            \label{fig:Contour_DK_Omicron}
        \end{figure}

        Figure \ref{fig:Contour_DK_Omicron} shows the contours of the true final size and the observed final size for $\rho = 0.5$. In both plots, the value of $a$ corresponding to the average daily tests carried out per citizen is highlighted, as well as the contours of observed final size of $45\%$ and $50\%$, approximately equal to the cumulative recorded cases of Omicron and all variants respectively. From the contour-plot of the true final size, figure \ref{fig:FinalSize_TrueContour_DK_Omicron}, we observe that the contours for the observed final size and the average testing effort correspond to a true final size between $65\%$ and $75\%$, in good correspondence with the serology study mentioned above.

\begin{figure}
        \begin{subfigure}[b]{\textwidth}
            \centering
            \includegraphics[width=\linewidth]{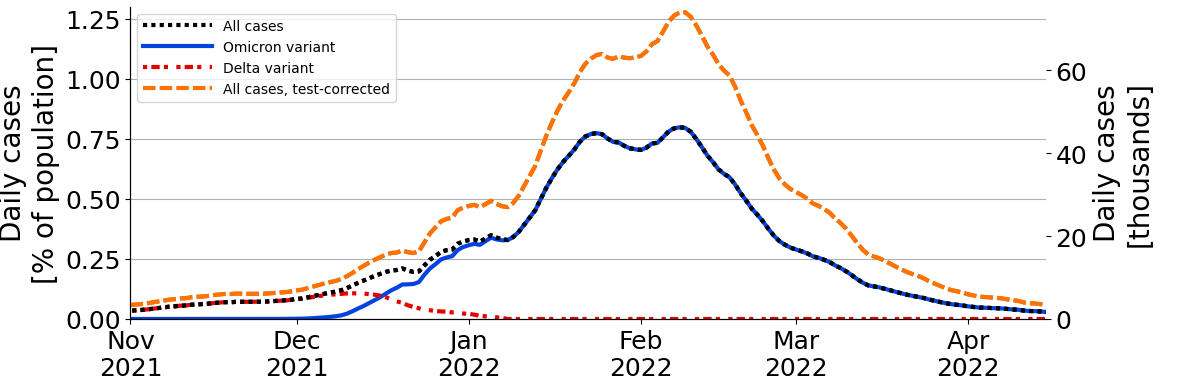}  
            \caption{7-day running mean of positive cases per day. The initial wave of Delta-variant infections is shown in red, while Omicron cases are shown in blue. The sum of all variants are shown in black. }
            \label{fig:DataCasesCorrected}
        \end{subfigure}
        \begin{subfigure}[b]{\textwidth}
            \centering
            \includegraphics[width=\linewidth]{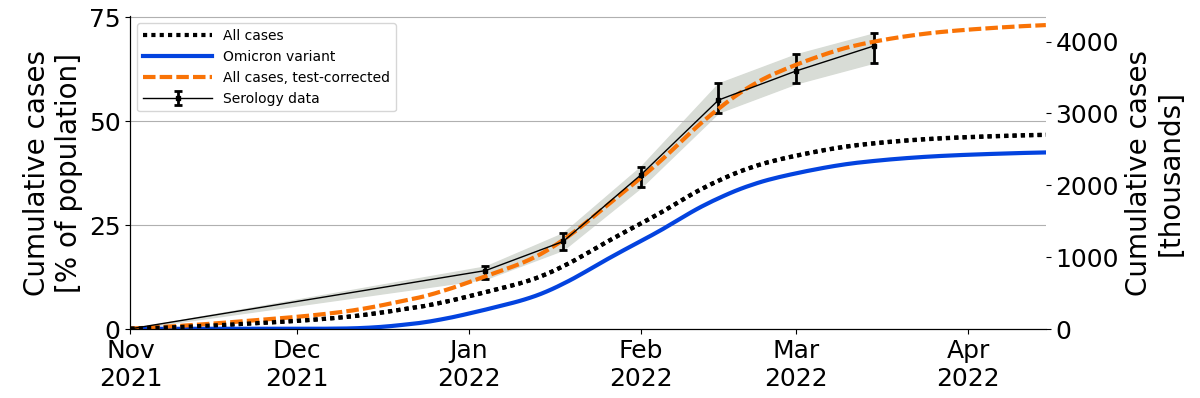}
            \caption{Cumulative cases of detected infections is shown in black, with cumulative Omicron-variant infections shown in blue. The estimate of population-wide serology from \citet{statens_serum_institut_ssi_seropraevalensundersogelse_2022} is shown as black bars with confidence intervals.}
            \label{fig:DataCasesCumulativeCorrected}
        \end{subfigure}
            \caption{}
            \label{fig:DataCorrectedSuper}
        \end{figure}

        For a more detailed analysis of the time-series data, we calculate the value of $\mathcal{A}$ on a daily basis, yielding a test-corrected estimate of the true incidence, i.e. $\mathcal{I}_{true} = \mathcal{I}_{observed}/\mathcal{A}$, where $\mathcal{I}_{true}$ are new infections and $\mathcal{I}_{observed}$ are the new infections which were recorded due to confirmatory and voluntary testing. 
        In figure \ref{fig:DataCorrectedSuper} we show the daily and cumulative time-series data, with the estimate of $\mathcal{I}_{true}$ based on testing-data. 

\clearpage
\pagebreak

\section{Discussion}




        In this work, we have presented a model extension of the classic SEIR-model in which symptom-drive (confirmatory) and voluntary testing were considered explicitly in the model structure. Distinguishing between infections that were identified and recorded and those that were not, allowed us to distinguish between the true size and the observed size of an epidemic. We determined expressions for the basic reproduction number, $\mathcal{R}_0$ and the ascertainment rate $\mathcal{A}$, i.e. the ratio between recorded infections and all infections, detailing how these quantities depend on model parameters. In particular, we investigated how the mentioned quantities relate to the proportion of infected individuals that get tested due to symptoms, $\rho$ and to the rate of voluntary testing of healthy and asymptomatic individuals, $\tau$. 

        Due to the structure of the model, the true epidemic size is always reduced when voluntary testing is increased, as individuals identified as infectious are assumed to quarantine and hence no longer contribute to future infections. 
        For some choices of parameters, the observed epidemic may however increase when voluntary testing is introduced since more cases are recorded. Further increases of the rate of voluntary testing may eventually decrease the observed epidemic size below the size observed in the absence of testing, since the true epidemic size will decrease with increased testing. 
        A consequence of this is that two epidemics waves of different sizes may be observed as equal in size due to differences in testing. As an example, we compared the SARS-CoV-2 pandemic waves in Denmark and Hungary at the end of 2020 and the beginning of 2021, for which the observed epidemic sizes were comparable. While the epidemics were observed to infect around $3\%$ of the population in both countries, differences in testing (both confirmatory and voluntary) suggests that, assuming everything equal, the true epidemic size in Hungary may have been twice that of the epidemic in Denmark.
        For small observed waves and low rates of voluntary testing, this is however very dependent on the parameter $\rho$ which represents the probability that symptomatic individuals get their infection confirmed by testing, which in turn depends on the availability of such confirmatory testing. 
        For $\rho$ close to one, the discrepancy between the true epidemic size and the observed epidemic size is low, since most of the recorded infections will be due to confirmatory testing. For low value of $\rho$ however, the feature that increasing voluntary testing also increases the observed epidemic is pronounced. In figure \ref{fig:AscertainmentContour_rhoR0} we illustrated how the observed epidemic will change when introducing some small amount of voluntary testing in a scenario without voluntary testing. The figure shows the boundary between two regimes: For low values of $\rho$ and higher values of $\mathcal{R}_0$, introducing voluntary testing will lead to an increase in the observed epidemic, while high values of $\rho$ and low values of $\mathcal{R}_0$ imply that introducing voluntary testing will cause the observed epidemic to decrease. 
        Hence, for diseases that spreads relatively fast (high $\mathcal{R}_0$) and where the onset (and confirmatory testing) of symptoms is rare (low $\rho$), the introduction of voluntary testing of asymptomatic individuals will cause an increase in observed incidence. If, however, the probability of symptomatic infections (and subsequent confirmatory testing) is high (high $\rho$), the introduction of voluntary testing with lead to a decrease in observed incidence, even for relatively high values of $\mathcal{R}_0$. Awareness of this difference between how voluntary testing affects observed incidence may be important when evaluating test-based strategies of mitigation of future epidemic threats. 

        We found that $1-\mathcal{A}$ scales with $(1+a)^{-3}$ where $a = \tau/\gamma$ is the rate of voluntary testing scaled by the rate of infection-progression. Similarly, the basic reproduction number $\mathcal{R}_0$ was found to scale with the sum of a factor of $(1+a)^{-2}$ and $(1+a)^{-3}$, with the former dominating. As an example, the former term is $\frac{16}{25} = 0.64$ for $a=\frac{1}{4}$, while it is $\frac{4}{9}\approx 0.44$ when $a=\frac{1}{2}$, suggesting that the reduction obtained from increasing $a$ by $\frac{1}{4}$ from $0$ reduced the contribution to new infections by $36\%$, while a further increase of $a$ by $\frac{1}{4}$ only reduced new infections by an additional $20\%$. These observations agree with the intuition that the reduction in infections due to voluntary testing diminishes as the rate of testing increases. Understanding how the reduction scales may be beneficial for decision-making regarding epidemic mitigation, as the decreasing benefit of increasing voluntary testing capacity affect the value of testing. 
        Similarly, the scaling of $1-\mathcal{A}$ implies that an initial increase of voluntary testing from zero provides more information about the true size of the epidemic than further increases of testing does. 
        In both cases, the scaling of $\mathcal{A}$ and $\mathcal{R}_0$ are direct consequences of the model structure, respectively due to the fact that we consider three stages at which voluntary testing can identify infected individuals and the fact that only the second and third stage of these (i.e. stage $P$ and $A$) contribute to new infections. As such, a model considering a different number of stages of infections would yield different results. The consequences of this and the appropriateness of choosing three stages requires additional investigation. 
        Alternatively, investigation into how different formulations of the process by which individuals progress through disease-stages may also be appropriate, e.g. a longer queued process more closely approximating a length of infection following Gamma-distribution. A similar challenge was discussed in our previous work on heterogeneity of testing \citep{berrig_heterogeneity_2022}. 
        Determining how the ascertainment rate scales with testing rates (both confirmatory and voluntary) is an important aim of future research, and requires more analysis of both the literature on mathematical modelling and on COVID-19 serology. While we have not presented a thorough review of these fields here, we briefly highlight one example, namely the work of \citet{macdonald_modelling_2021}, also mentioned in the introduction. The authors argue the ascertainment rate scales testing in such a way that $\mathcal{A} = \frac{k\tau}{A + \tau} + k_0$ where $k$, $k_0$ and $A$ are model parameters, a functional form not too dissimilar from the expression found in the present work. The work of Macdonald and colleagues are however concerned with the first epidemic wave of spring 2020, where most testing was assumed to be for confirmation of systems, and where testing efforts were still low. Detailed comparison of this work, and other similar works, may be the key to understand how the ascertainment rate varied over different phases of the COVID-19 pandemic.


        While much of the analysis discussed in the present work relates to the sum of infections following an epidemic, we also applied our results the time-series data of testing during the Omicron wave in Denmark in the winter of 2021/2022, as discussed at the end of section \ref{sec:Omicron}. 
        As an estimate for the rate of voluntary testing, we considered the daily sum of negative PCR- and LFT-tests divided by the number of individuals that had not yet been found positive (under the assumption that most individuals abstained from voluntary testing in the immediate period after being found positive). Assuming that $\rho=0.5$ and $\gamma = 1/3$, we calculated a daily value of $\mathcal{A}$ and used it to correct daily incidence-data and obtain an estimate for the true number of new daily infections. The results, shown in figure \ref{fig:DataCorrectedSuper} showed a very good agreement with serology data from the Omicron wave, while also providing time-series for the true epidemic wave. Interestingly, the two peaks of equal size seen in late January and mid February 2022 in the observed incidence data, are revealed to be of different sizes when correcting for the decrease in testing during this period. 
        While improved estimation of the $\rho$ and $\gamma$ parameters could lead to a refined expression for the numerical relation between data and true incidence, our results give a method for correcting incidence data based on mechanistic considerations about disease dynamics. 

        In our work, we do not distinguish between the types of tests used, despite the two types of test primarily used during the Omicron wave in Denmark (PCR and LFT) having different sensitivity and specificity. These differences could be built into the model. The increased sensitivity of PCR-testing could be modelled by allowing individuals in the exposed stage, $E$, to be found positive. Furthermore, testing rates could be scaled according to disease stage, reflecting the sensitivity of different test-types during specific disease stages. 
        The low specificity of LFT implies an increased risk of false positive tests, which may lead to unnecessary quarantining of healthy individuals. While LFT-positive individuals were advised to get their infection confirmed by PCR-testing, the quarantining of healthy but susceptible individuals may also have had an effect on epidemic dynamics. An extension of the model in which temporary quarantining of susceptible individuals also occurs at a rate proportional to the rate of voluntary testing would more accurately capture the effects of high rates of voluntary testing with low-specificity tests. While such investigations will be explored in the future, we expect the effect on the ascertainment rate to be minor.

        An important difference between the different types of tests used during the COVID-19 pandemic was the delay between when the test was carried out, and when the result of the test was available. The results of PCR-tests were typically not available to the tested individual until a day following testing, while LFTs provided near immediate results, typically within 20-30 minutes. A consequence is that exposed individuals tested by PCR-test may have been spreading infections even after having had a test that eventually returns a positive result. 
        This was one of the key points discussed by \citet{larremore_test_2020} before wide-scale testing with LFTs were initiated in multiple countries, and may have a large impact on the reduction in disease spread achieved through testing. 
        In the work discussed in this paper, we did not consider such delay in test-results. While this could be incorporated in the model, e.g. as a delay-dependent reduction of $\mathcal{R}_0$ as done by \citet{berrig_heterogeneity_2022}, we here focused on the effect that testing had on the ascertainment ratio $\mathcal{A}$ rather than the effect that testing had on mitigation. 
        In future work, we plan to study in greater detail how delays in test-results may affect large-scale testing as a mitigation strategy.         

        High rates of testing were in many countries combined with contact-tracing efforts in which close contacts of identified cases were traced and instructed to isolate and test. The potential for contact-tracing to improve the epidemic mitigation achieved through testing may be significant, and has been explored thoroughly by multiple authors, including those already highlighted in the introduction \citep{heidecke_mathematical_2024,zhang_analysing_2022,sturniolo_testing_2021,barbarossa_fleeing_2021,kretzschmar_impact_2020}. 
        In our work, we did not include contact-tracing, in order to reduce the number of model assumptions and parameters. While combining our model with many of the contact-tracing models discussed in the literature would be feasible, the varying orders of magnitude of voluntary testing and infections observed in Denmark, particularly during the Omicron wave, means that contact-tracing efforts cannot be assumed to have been constant during the pandemic. In periods where testing rates were high, but incidence was low, tracing potential infections required little effort, but in the beginning of 2022 when almost one in four citizens tested positive within a single month, such efforts may have been entirely futile. 
        However, the reduction that contact-tracing may provide in the force of infection, and consequently in the reproduction number, means that the benefits of voluntary testing may have been underestimated in this work. Particularly the difference between a low rate of testing and no testing (neither confirmatory nor voluntary) may be very different when efficient contact-tracing is applied.

        Differences in testing behavior has lead to major communicative challenges across both time and country borders. At the end of the Omicron wave, Denmark experienced a large increase in deaths registered as COVID-19-related. In the highly vaccinated population, it was likely that this increase was only to a little degree due to infection-related fatalities, and rather due to deaths from other sources that had occurred following a positive COVID-19 test, as suggested by considerations about general mortality and deaths certificates \citep{friis_covid-19_2023}. As definitions of COVID-19 deaths rely on the testing status of individuals, it is necessary to correct for testing efforts and estimate the true epidemic wave accurately. 

        The results presented in this work provide a method for correcting incidence data for high rates of voluntary testing. In conjunction with detailed serology data, excess mortality calculations and other methodologies, the presented results may improve surveillance of epidemic diseases and allow for better comparison between countries during both the COVID-19 pandemic and future pandemics.

\clearpage
\pagebreak 

\bibliography{CovidTestModelling}

\clearpage
\pagebreak 

\appendix 
\section{Supplementary material}

\subsection{Analysis of the general form of the model}\label{sec:AnalysisGeneral}

        In the presentation of the model, all disease-progression parameters were introduced with a subscript related to the specific infection-stage it related to. 
        For most of the analysis however, we assumed all pre-symptomatic stages to progress with the same rate, $\gamma$.
        For completeness, we here show the expression for the reproduction number, $\mathcal{R}_0$ and the ascertainment ratio, $\mathcal{A}$, in the most general form of the model, maintaining subscripts of all disease-progression parameters.

        The basic reproduction number is given by
        \begin{equation}
            \mathcal{R}_0 = \beta \left( \dfrac{\gamma_L}{(\gamma_L + \tau_L)(\gamma_P + \tau_P)} + \dfrac{\gamma_L \gamma_P ( 1-\rho)}{(\gamma_L + \tau_L)(\gamma_P + \tau_P)(\gamma_A+\tau_A)}\right).
            \label{eq:R0_general}
        \end{equation}


        The ascertainment ratio can be written as
        \begin{equation}
            \mathcal{A} = 1 - \dfrac{\gamma_L \gamma_P \gamma_A( 1-\rho)}{(\gamma_L + \tau_L)(\gamma_P + \tau_P)(\gamma_A+\tau_A)}
        \end{equation}
        or, equivalently:
        \begin{equation}
            \mathcal{A} = \dfrac{\tau_L}{\gamma_L + \tau_L} + \dfrac{\gamma_L(\tau_P + \gamma_P\rho)}{(\gamma_L+\tau_L)(\gamma_P+\tau_P)} + \dfrac{\tau_A \gamma_L \gamma_P( 1-\rho)}{(\gamma_L + \tau_L)(\gamma_P + \tau_P)(\gamma_A+\tau_A)}
        \end{equation}

\subsection{Next Generation Matrix with large domain}\label{sec:NGMlarge}
            For completeness, we here give the full Next Generation Matrix with large domain, as defined by \citep{diekmann_construction_2010}.
            The infected subsystem consists of variables $(E,L,P,A)$ since only these may contribute to future infections. 
            \begin{equation}
                \bm{T} = \begin{pmatrix}
                    0 & 0 & \beta S & \beta S \\
                    0 & 0 & 0 & 0 \\
                    0 & 0 & 0 & 0 \\
                    0 & 0 & 0 & 0 \\
                \end{pmatrix}
            \end{equation} and 
            \begingroup
            \renewcommand*{\arraystretch}{2.2}
            \begin{equation}
                \bm{\Sigma} = \begin{pmatrix}
                    \gamma & 0 & 0 & 0\\
                    -\gamma & \gamma+\tau & 0 & 0\\
                    0 & -\gamma & \gamma+\tau & 0 \\
                    0 & 0 & -\gamma(1-\rho) & \gamma + \tau\\
                \end{pmatrix}\label{eq:MatrixVLarge}
            \end{equation}
            \endgroup
            
            The inverse of $\bm{\Sigma}$ is trivially computed as:            
            \begingroup
            \renewcommand*{\arraystretch}{2.2}
            \begin{equation}
                \bm{\Sigma}^{-1} = \begin{pmatrix}
                    \dfrac{1}{\gamma} & 0 & 0 & 0\\
                    \dfrac{1}{\gamma+\tau} & \dfrac{1}{\gamma+\tau} & 0 & 0\\
                    \dfrac{\gamma}{(\gamma+\tau)^2} & \dfrac{\gamma}{(\gamma+\tau)^2} & \dfrac{1}{\gamma+\tau} & 0 \\
                    \dfrac{\gamma^2 (1-\rho)}{(\gamma + \tau)^3} & \dfrac{\gamma^2 (1-\rho)}{(\gamma + \tau)^3} & \dfrac{\gamma (1-\rho)}{(\gamma + \tau)^2} & \dfrac{1}{\gamma + \tau}  \\ 
                \end{pmatrix}\label{eq:MatrixVInverseLarge}
            \end{equation}
            \endgroup

\subsection{Reduced form of the model}\label{sec:ReducedForm}

        We here present a reduced form of the suggested mathematical model.
        For the reduced form, we assume that all rates of disease-progression are the same, $\gamma = \gamma_E = \gamma_L = \gamma_P = \gamma_A = \gamma_I = \gamma_Q$. We introduce a scaling of time such that $T = \gamma t$. Denoting $\frac{\partial X}{\partial T} = X'$ and defining $a = \tau/\gamma$ and $b = \beta/\gamma$, the model system can be written as:

        \begin{align}
            S' &= - b S(P+A) \\
            E' &= b S (P+A) - E \\
            L' &= E - (1+a)L \\
            P' &= L - (1+a)P \\
            A' &= (1-\rho) P - (1+a)A \\
            I' &= \rho P - aI \\
            Q' &= a (L+P+A) - Q \\
            R_p' &= Q + I \\
            R_n' &= A
        \end{align}

        Using the same methods as described in the main text, we find that the basic reproduction number near the DFE where $S=1$ is given by
        \begin{equation}
            \mathcal{R}_0 = b \left(\dfrac{1}{(1+a)^2} + \dfrac{1-\rho}{(1+a)^3}\right) 
        \end{equation}
        and that the ascertainment ratio is given by
        \begin{equation}
            \mathcal{A} = 1-\dfrac{1-\rho}{(1+a)^3}
        \end{equation}

\subsection{Final Size Calculations}\label{sec:FinalSizeCalculations}




        As $t\rightarrow \infty$, the model system approaches a steady state without any active cases.
        We follow the methodology previously considered by \citet{andreasen_epidemics_2018}. 

        For notational purposes, we define for each variable $y$, the integral over the full epidemic as $T_y = \int_0^{\infty} y dt$. 

        From the system of differential equations given in equations \eqref{eq:modeldefinition}, we write up the following quantities:
        \begin{subequations}
        \begin{align}
            \dot{S} / S &= - \beta (P + A) \\
            \dot{S} + \dot{E} + \dot{L} &= -(\gamma + \tau) L \\
            \dot{S} + \dot{E} + \dot{L} +\dot{P} &= -(\gamma + \tau) P - \tau L \\
            \dot{S} + \dot{E} + \dot{L} + \dot{P} + \dot{A} &= -(\gamma + \tau) A -(\gamma \rho + \tau) P - \tau L 
        \end{align}\label{eq:FinalCalcInit}
        \end{subequations}

        As $t$ approaches infinity, the stability of the systems implies that all variables apart from $S$, $R_p$ and $R_n$ are zero. We denote that final size of these variables as $S(t) \underset{t\rightarrow \infty}{\longrightarrow} \sigma$, $R_p(t) \underset{t\rightarrow \infty}{\longrightarrow} r_p$ and $R_n(t) \underset{t\rightarrow \infty}{\longrightarrow} r_n$

        Integrating equations \eqref{eq:FinalCalcInit} from $t=0$ to $t=\infty$ yields:
        \begin{subequations}
        \begin{align}
            \log{\sigma} &= - \beta (T_P - T_A) \\
            \sigma - S_0 - E_{0} - L_{0} &= - (\gamma + \tau) T_{L} \\
            \sigma - S_0 - E_{0} - L_{0} - P_0&= -(\gamma + \tau) T_P - \tau T_{L} \\
            \sigma - S_0 - E_{0} - L_{0} - P_0 - A_0&= -(\gamma + \tau) T_A -(\gamma \rho + \tau) T_P - \tau T_{L} 
        \end{align}\label{eq:FinalCalcInt}
        \end{subequations}
        Where $Y_0$ denote the initial condition for variable $Y$. 

        Furthermore, observe that the equations for $\dot{R}_n$ and $\dot{R}_p$, equations \eqref{eq:modeldefinitionRn} and \eqref{eq:modeldefinitionRp} respectively, when integrated from $t=0$ to $t=\infty$ yields:

        \begin{align}
            r_p - R_{p,0} &= \gamma T_Q + \gamma T_I \label{eq:rpinfty} \\
            r_n - R_{n,0} &= \gamma T_A \label{eq:rninfty}
        \end{align}






        In general, we consider initial conditions such that the vast majority of the population is initially susceptible, $S_0\approx 1$, and the initial number of cases is low, $0 < E_{0} \ll 1$.
        In the limit where $S_0 \rightarrow 1$, with $E_{0} \rightarrow 0$, $L_{0} \rightarrow 0$, $P_0 \rightarrow 0$ and $A_0 \rightarrow 0$, equations \eqref{eq:FinalCalcInt} become:
        \begin{subequations}
        \begin{align}
            \log{\sigma} &= - \beta (T_P - T_A) \label{eq:FinalCalcIsolatedLogSigma} \\ 
            \sigma &= 1 - (\gamma + \tau) T_{L} \\
            \sigma &= 1 -(\gamma + \tau) T_P - \tau T_{L} \\
            \sigma &= 1 -(\gamma + \tau) T_A -(\gamma \rho + \tau) T_P - \tau T_{L} 
        \end{align}\label{eq:FinalCalcS1}
        \end{subequations}

        Assuming $T_P + T_A \neq 0$, this can be written as:
        \begin{subequations}
        \begin{align}
            \beta &= \dfrac{-\log \sigma}{T_P + T_A} \label{eq:FinalCalcIsolatedBeta} \\
            T_{L} &= \frac{1}{\gamma + \tau} \left(1-\sigma \right)  \\
            T_P &= \frac{1}{\gamma + \tau}\left( 1 - \sigma - \tau T_{L} \right)\\
            T_A &= \frac{1}{\gamma + \tau} \left(1 - \sigma -(\gamma \rho + \tau) T_P - \tau T_{L}\right)
        \end{align}\label{eq:FinalCalcIsolated}
        \end{subequations}

        We note that equation \eqref{eq:FinalCalcIsolatedBeta} describes a relation between $\beta$ and $\sigma$. Since $T_P$ and $T_A$ are described in terms of $\gamma$, $\tau$, and $\sigma$, it is possible to use equation \eqref{eq:FinalCalcIsolatedBeta} to determine a value of $\beta$ that yields a particular $\sigma$. 

        Replacing $X = 1 - \sigma$ we find:

        \begin{subequations}
        \begin{align}
            \beta &= \dfrac{-\log (1-X)}{T_P + T_A} \label{eq:FinalCalcIsolatedBeta2} \\
            T_{L} &= \frac{1}{\gamma + \tau} X  \\
            T_P &= \frac{1}{\gamma + \tau}\left( X - \tau T_{L} \right) = \dfrac{\gamma}{(\gamma+\tau)^2} X \\
            T_A &= \frac{1}{\gamma + \tau} \left(X -(\gamma \rho + \tau) T_P - \tau T_{L}\right) = \dfrac{\gamma^2 (1-\rho)}{(\gamma + \tau)^3} X
        \end{align}\label{eq:FinalCalcIsolated2}
        \end{subequations}

        Plugging $T_A$ and $T_P$ into \eqref{eq:FinalCalcIsolatedBeta2} yields the expression
        \begin{align}
            \log(1-X) &= - \beta \left(\dfrac{\gamma}{(\gamma+\tau)^2} + \dfrac{\gamma^2 (1-\rho)}{(\gamma + \tau)^3}) \right) X \\
            \log(1-X) &= - \mathcal{R}_0 X  \\
            X & = 1 - \exp(-\mathcal{R}_0 X)
        \end{align}
        with $\mathcal{R}_0$ as defined in equation \eqref{eq:R0} of the main text.

\subsection{Model extension when relaxing confirmatory testing assumption}\label{sec:ModelExtensionY}

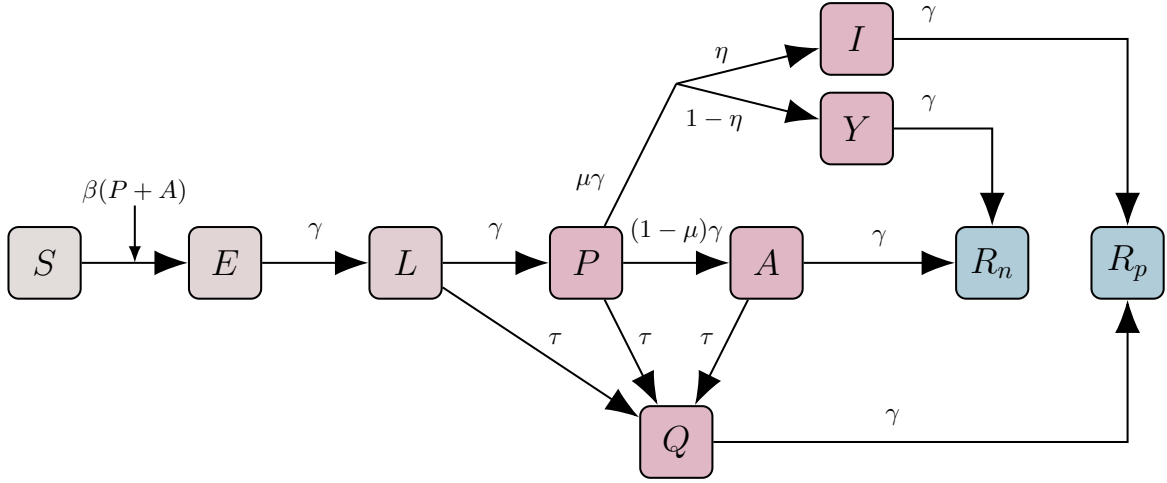
\begin{figure}[ht]
            \centering
            \scalebox{0.95}{
                \begin{tikzpicture}[thick,scale=1, every node/.style={scale=1}]
                    \node[rectangle,rounded corners,fill=susColor,draw,minimum size =  \CompSize cm] (S) at (0,0) {\Large $S$};
                    \node[rectangle,rounded corners,fill=susColor!80!infColor,draw,minimum size =  \CompSize cm] (E1) at (\cd,0) {\Large $E$};
                    \node[rectangle,rounded corners,fill=susColor!60!infColor,draw,minimum size =  \CompSize cm] (E2) at (2*\cd,0) {\Large $L$};
                    \node[rectangle,rounded corners,fill=infColor,draw,minimum size =  \CompSize cm] (P) at (3*\cd,0)  {\Large $P$};
                    \node[rectangle,rounded corners,fill=infColor,draw,minimum size =  \CompSize cm] (A) at (4*\cd,0) {\Large $A$};
                    \node[rectangle,rounded corners,fill=infColor,draw,minimum size =  \CompSize cm] (Q) at (3.5*\cd,-\cd)  {\Large $Q$};
                    
                    \node (Iold) at (3.5*\cd,\cd) {};
                    \node[rectangle,rounded corners,fill=infColor,draw,minimum size =  \CompSize cm] (I) at (4.5*\cd,1.25*\cd) {\Large $I$};
                    \node[rectangle,rounded corners,fill=infColor,draw,minimum size =  \CompSize cm] (Y) at (4.5*\cd,0.75*\cd) {\Large $Y$};
                    
                    \node[rectangle,rounded corners,fill=recColor,draw,minimum size =  \CompSize cm] (Rn) at (5.25*\cd,0) {\Large $R_n$};
                    \node[rectangle,rounded corners,fill=recColor,draw,minimum size =  \CompSize cm] (Rp) at (6*\cd,0) {\Large $R_p$};

                    \draw [-{Latex[scale=2]}]  (S) -- (E1);
                    \draw [-{Latex[scale=2]}]  (E1) -- (E2);
                    \draw [-{Latex[scale=2]}]  (E2) -- (P);
                    \draw [-{Latex[scale=2]}]  (P) -- (A);
                    
                    
                    \draw [-]  (P) -- (Iold.center);
                    \draw [-{Latex[scale=2]}]  (Iold.center) -- (I);
                    \draw [-{Latex[scale=2]}]  (Iold.center) -- (Y);
                    
                    \draw [-{Latex[scale=2]}]  (E2) -- (Q);
                    \draw [-{Latex[scale=2]}]  (P) -- (Q);
                    \draw [-{Latex[scale=2]}]  (A) -- (Q);
                    \draw [-{Latex[scale=2]}]  (A) -- (Rn);
                    \draw [-{Latex[scale=2]}]  (Q) -- ($(Rp)+(0,-\cd)$) -- (Rp);
                    \draw [-{Latex[scale=2]}]  (I) -- ($(Rp)+(0,1.25*\cd)$) -- (Rp);
                    \draw [-{Latex[scale=2]}]  (Y) -- ($(Rn)+(0,0.75*\cd)$) -- (Rn);

                    \draw [-{Latex[scale=1]}]  ($(S)!0.5!(E1) + (0,0.8)$) -- ($(S)!0.5!(E1)$);
                    \node at ($(S)!0.5!(E1) + (0,1)$) {$\beta (P+A)$};

                    \node at ($(E1)!0.5!(E2)+(0,0.45)$) {$\gamma$};
                    \node at ($(E2)!0.5!(P)+(0,0.45)$) {$\gamma$};
                    \node at ($(P)!0.5!(A)+(0,0.45)$) {$(1-\mu)\gamma$};
                    \node at ($(P)!0.25!(Iold)+(-0.25,0.5)$) {$\mu \gamma$};
                    \node at ($(Iold)!0.25!(I)+(0,0.25)$) {$\eta$};
                    \node at ($(Iold)!0.25!(Y)+(-0.1,-0.35)$) {$1-\eta$};
                    \node at ($(E2)!0.5!(Q)+(0.2,0.2)$) {$\tau$};
                    \node at ($(P)!0.5!(Q)+(0.2,0.2)$) {$\tau$};
                    \node at ($(A)!0.5!(Q)+(-0.2,0.2)$) {$\tau$};
                    \node at ($(A)!0.5!(Rn)+(0,0.35)$) {$\gamma$};
                    \node at ($(I)+(1,0.35)$) {$\gamma$};
                    \node at ($(Y)+(1,0.35)$) {$\gamma$};
                    \node at ($(Q)+(3,0.35)$) {$\gamma$};

                \end{tikzpicture}
            }
            \caption{Compartment diagram of the model when relaxing the assumption that confirmatory tests are carried out for all symptomatic individuals. }
            \label{fig:ModelFigureY}
        \end{figure}

        In the model presented in the main text, we assume that all symptomatic individuals ($I$) are tested in order to confirm their infection. Relaxing this assumption, we can distinguish between symptomatic individuals that confirm their infection by testing (reusing notation, we denote this group $I$) and those that do not carry out confirmatory testing.
        We denote the proportion of infection individuals that develop symptoms as $\mu$ (for the main model this was denoted $\rho$), and the fraction that confirm the infection by $\eta$. 
        In figure \ref{fig:ModelFigureY} a compartment diagram of this model is shown, while the model-equations for $A$, $I$, $Y$, and $R_n$ are given in equations \eqref{eq:ModelExtension}. The remaining model-equations are unchanged from those given in equation \eqref{eq:modeldefinition} of the main text.

        \begin{subequations}
            \begin{align}
                \dot{A} &= \gamma (1-\mu) P - (\gamma + \tau) A \\
                \dot{I} &= \gamma \mu \eta P - \gamma I \\
                \dot{Y} &= \gamma \mu (1-\eta) P - \gamma Y \\
                \dot{R_n} &= \gamma A + \gamma Y
            \end{align}\label{eq:ModelExtension}
        \end{subequations}

        We assume that individuals in the $Y$ compartment also quarantine, and hence they do not contribute to the force of infection, leaving the calculation of $\mathcal{R}_0$ unchanged (albeit with $\mu$ replacing $\rho$).

        Through similar calculations as those given in the main text, the ascertainment ratio can be shown to be
        \begin{equation}
            1-\mathcal{A}_{\eta} = \dfrac{\gamma^2}{(\gamma + \tau)^2} \left(\dfrac{\gamma (1-\mu)}{\gamma+\tau} + \mu(1-\eta)\right) =\dfrac{\gamma^3 (1-\mu)}{(\gamma + \tau)^3} + \dfrac{\gamma^2 \mu(1-\eta) }{(\gamma + \tau)^2}
        \end{equation}
        where the subscript $\eta$ is used to distinguish from the ascertainment ratio of the main model. Hence, relaxing the assumption contributes a factor of $-\frac{\gamma^2 \mu(1-\eta) }{(\gamma + \tau)^2}$ to $\mathcal{A}_{\eta}$. 
        Note that for $\tau=0$, the ascertainment ratio is given by $\mathcal{A}_{\eta} = 1 - \mu \eta$. For low $\tau$, the extended model can be approximated by the main model by setting $\rho = \mu \eta$. 

        In reduced time-units (see supplementary section \ref{sec:ReducedForm}), the ascertainment ratio can be written as 
        \begin{equation}
            1-\mathcal{A}_{\eta} = \dfrac{1-\mu}{(1+a)^3} + \dfrac{\mu(1-\eta)}{(1+a)^2}
        \end{equation}
        or, alternatively, as 
        \begin{equation}
            \mathcal{A}_{\eta} = \dfrac{a}{1+a} + \dfrac{a+\mu\eta}{(1+a)^2} + \dfrac{a(1-\mu)}{(1+a)^3}
        \end{equation}

\begin{figure}[!ht]
            \centering
            \includegraphics[width= 0.8 \linewidth]{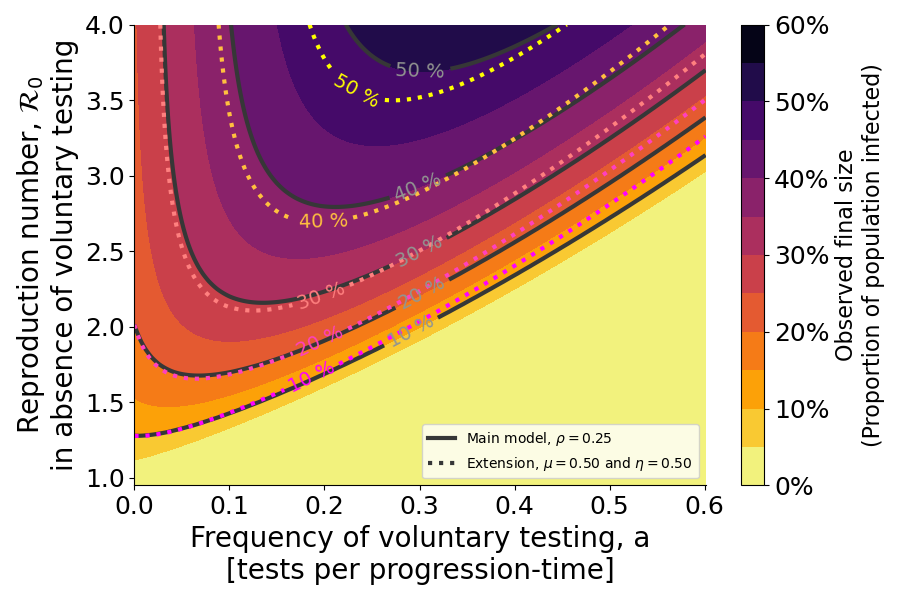}
            \caption{Contour-plot of the observed final size for both the extended model, shown as dotted lines, and the model from the main text, shown as filled contours and full black lines. In the extended model, $\mu = 0.5$ and $\eta = 0.5$, while the main model has the comparable choice of parameters of $\rho = \mu \eta = 0.25$.}
            \label{fig:ContourExtensionComparison}
        \end{figure}

\begin{figure}[!ht]
            \centering
            \includegraphics[width= 0.8 \linewidth]{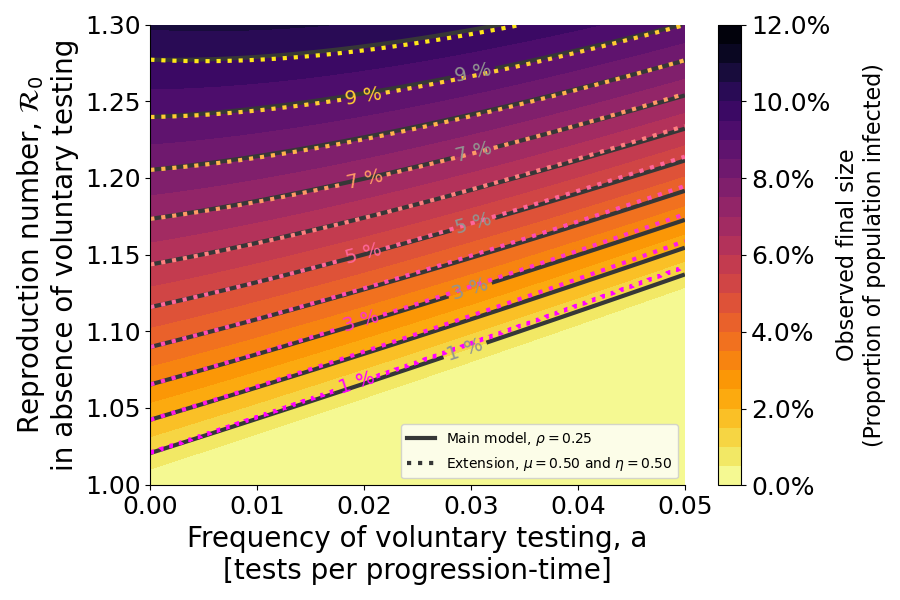}
            \caption{Contour-plot of the observed final size for both the extended model, shown as dotted lines, and the model from the main text, shown as filled contours and full black lines. In the extended model, $\mu = 0.5$ and $\eta = 0.5$, while the main model has the comparable choice of parameters of $\rho = \mu \eta = 0.25$.}
            \label{fig:ContourExtensionComparisonZoom}
        \end{figure}

        In figure \ref{fig:ContourExtensionComparison} an example is shown of how the extension changes the relation between voluntary testing and the observed final size, with figure \ref{fig:ContourExtensionComparisonZoom} showing the same contours, zoomed in on a regime with a smaller epidemic. We observe that for comparison where $\rho = \mu \eta$, the two models give comparable results. The differences increase for lower values of $\rho = \mu\eta$ and diminish for higher values (not shown). However, we find that despite the additional dynamics captured by extension, the simplicity of the main model is preferable. In the main text, we focus on the main model, and interpret the value of $\rho$ as an approximation for $\mu \eta$, i.e. the product of the ratio of infected developing symptoms and the probability that a confirmatory test is carried out.




\clearpage

\end{document}